\pdfoutput=1
\documentclass[a4,prb,twocolumn,amsfonts,amssymb,showpacs]{revtex4-1}
\usepackage{amsmath}
\usepackage{graphicx}
\usepackage{color}
\usepackage{hyperref}
\usepackage{psfrag}
\usepackage{stmaryrd}
\usepackage{amssymb}
\usepackage{wasysym}

\begin{document}

\title{Viscous spreading of an inertial wave beam in a rotating fluid}

\author{Pierre-Philippe Cortet}
\email{ppcortet@fast.u-psud.fr}
\author{Cyril Lamriben}
\author{Fr\'{e}d\'{e}ric Moisy}

\affiliation{Laboratoire FAST, CNRS UMR 7608, Universit\'{e}
Paris-Sud, Universit\'{e} Pierre-et-Marie-Curie, B\^{a}t. 502,
Campus universitaire, 91405 Orsay, France}

\date{\today}

\begin{abstract}

We report experimental measurements of inertial waves generated by
an oscillating cylinder in a rotating fluid. The two-dimensional
wave takes place in a stationary cross-shaped wavepacket. Velocity
and vorticity fields in a vertical plane normal to the wavemaker
are measured by a corotating Particule Image Velocimetry system.
The viscous spreading of the wave beam and the associated decay of
the velocity and vorticity envelopes are characterized. They are
found in good agreement with the similarity solution of a linear
viscous theory, derived under a quasi-parallel assumption similar
to the classical analysis of Thomas and Stevenson [J. Fluid Mech.
{\bf 54} (3), 495--506 (1972)] for internal waves.

\end{abstract}

\maketitle

\section{Introduction}

Rotating and stratified fluids both support the propagation of
waves, referred to as inertial and internal waves respectively,
which share numbers of similar
properties.\cite{Greenspan1968,Lighthill1978} These waves are of
first importance in the dynamics of the ocean and the
atmosphere,\cite{Pedlosky1987} and play a key role in the
anisotropic energy transfers and in the resulting quasi-2D nature
of turbulence under strong rotation and/or
stratification.\cite{Cambon2001}

More specifically, rotation and stratification (and the
combination of the two) lead to an anisotropic dispersion relation
in the form $\sigma = f(k_z / |{\bf k}|)$, where $\sigma$ is the
pulsation, ${\bf k}$ is the wave vector, and the $z$ axis is
defined either by the rotation axis or the
gravity.\cite{Lighthill1978} This particular form implies that a
given excitation frequency $\sigma$ selects a single direction of
propagation, whereas the range of excited wavelengths is set by
boundary conditions or viscous effects. A number of well-known
properties follow from such dispersion relation, such as
perpendicular phase velocity and group velocity, and anomalous
reflection on solid boundaries.\cite{Lighthill1978,Phillips1963}

Most of the laboratory experiments on internal waves in stratified
fluids have focused on the properties of localized wave beams, of
characteristic thickness and wavelength much smaller than the size
of the container, excited either from
local\cite{Mowbray1967,Thomas1972,Sutherland1999,Flynn2003,Gostiaux2006}or
extended\cite{Gostiaux2007} sources. On the other hand, most of
the experiments in rotating fluids have focused on the inertial
modes or wave attractors in closed
containers,\cite{Fultz1959,McEwan1970,Manasseh1994,Maas2001,Meunier2008}
whereas less attention has been paid to localized inertial wave
beams in effectively unbounded systems. Inertial modes and
attractors are generated either from a disturbance of significant
size compared to the container,\cite{Fultz1959} or more
classically from global forcing (precession or modulated angular
velocity).\cite{McEwan1970,Manasseh1994,Maas2001,Meunier2008}
Localized inertial waves generated by a small disturbance have
been visualized from numerical simulations by Godeferd and
Lollini,\cite{Godeferd1999} and have been recently investigated
using Particle Image Velocimetry (PIV) by  Messio \textit{et
al}.\cite{Messio2008} In this latter experiment, the geometrical
properties of the conical wavepacket emitted from a small
oscillating disk was characterized, by means of velocity
measurements restricted to a horizontal plane normal to the
rotation axis, intersecting the wavepacket along an annulus.

The weaker influence of rotation compared to stratification in
most geophysical applications probably explains the limited number
of references on inertial waves compared to the abundant
literature on internal waves (see Voisin \cite{Voisin2003} and
references therein). Another reason might be that quantitative
laboratory experiments on rotating fluids are more delicate to
perform than for stratified fluids. In particular, the velocity
field of internal waves is 2 components, whereas that of inertial
waves is 3 components (although the geometry of the wave pattern
may be two-dimensional). Moreover, only PIV is available for
quantitative investigation of the wave structure for inertial
waves, whereas other optical methods, such as shadowgraphy, or
more recently synthetic Schlieren,\cite{Sutherland1999} are also
possible for internal waves.

The purpose of this paper is to extend the results of Messio
\textit{et al.},\cite{Messio2008} using a newly designed rotating
turntable, in which the velocity field can be measured over a
large vertical field of view using a corotating PIV system. In the
present experiment, the inertial wave is generated by a thin
cylindrical wavemaker, producing a two-dimensional cross-shaped
wave beam, and special attention is paid to the viscous spreading
of the wave beam. The beam thickness and the vorticity decay are
found to compare well with a similarity solution, analogous to the
one derived by Thomas and Stevenson\cite{Thomas1972} for internal
waves.

\section{Theoretical background}

\subsection{Geometry of the wave pattern}

\begin{figure}
\psfrag{x}[l][][0.8]{$x$} \psfrag{z}[l][][0.8]{$z$}
\psfrag{X}[l][][0.8]{$X$}\psfrag{Y}[l][][0.8]{$Y$}
\psfrag{Z}[l][][0.8]{$Z$} \psfrag{O}[c][][0.8]{${\bf \Omega}$}
\psfrag{vp}[c][][0.8]{${\bf c}$} \psfrag{vg}[c][][0.8]{${\bf
c}_g$} \psfrag{k}[c][][0.8]{${\bf k}$}\psfrag{u}[c][][0.8]{${\bf
u}$} \psfrag{T}[c][][0.8]{$\theta$} \psfrag{vx}[c][][0.8]{$u_x$}
\psfrag{as}[r][][0.7]{$Z_0(t)$}
\psfrag{d}[c][][0.8]{$\delta$}\psfrag{2R}[c][][0.8]{$2R$}
\centerline{\includegraphics[width=8cm]{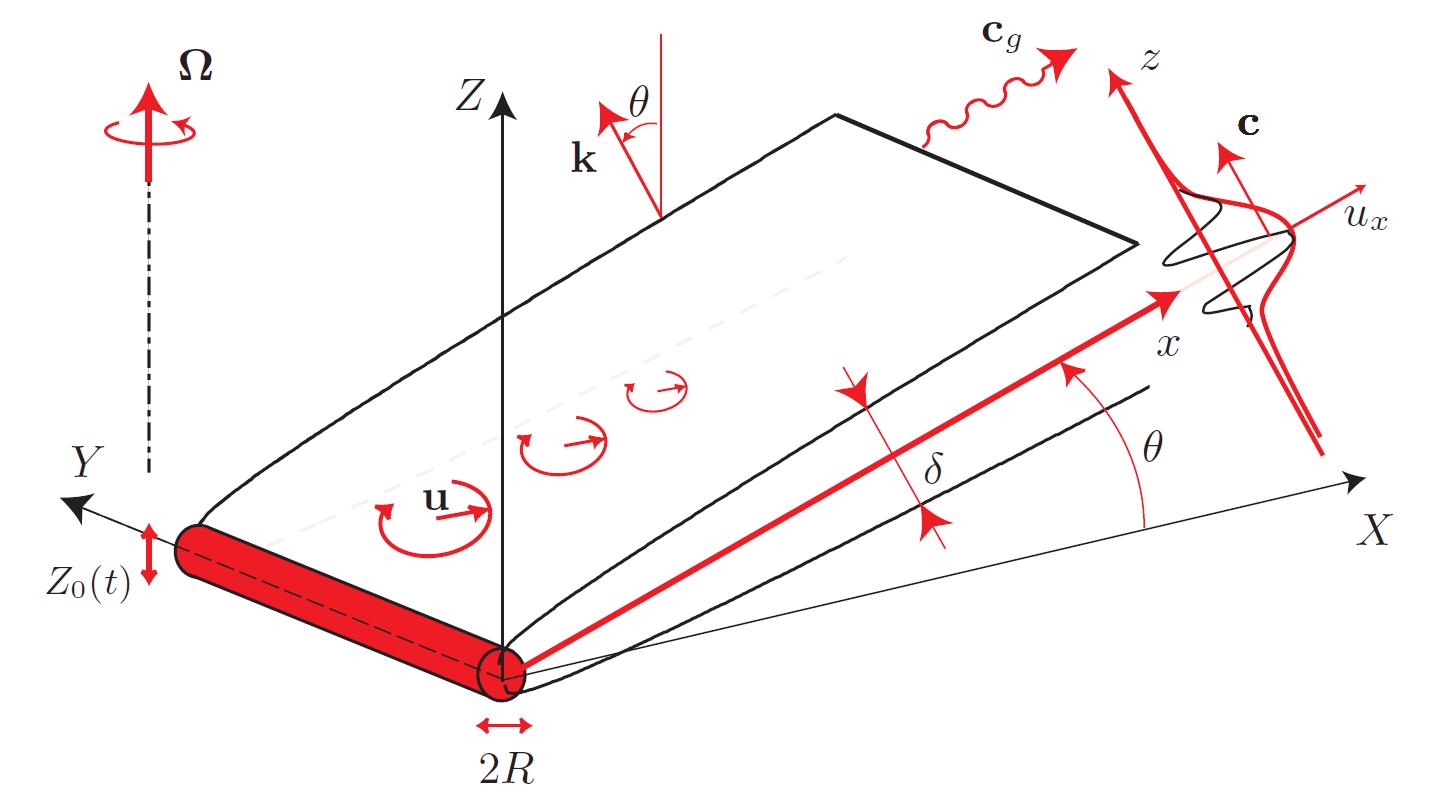}}
\caption{Geometry of an inertial wave beam emitted in an infinite
medium from a localized oscillating cylindrical wavemaker
invariant in the $Y$-direction.} \label{fig:frame}
\end{figure}

A detailed description of the structure of a plane monochromatic
inertial wave in an inviscid fluid can be found in Messio
\textit{et al.},\cite{Messio2008} and only the main properties are
recalled here.  We consider a fluid rotating at constant angular
velocity ${\boldsymbol \Omega} = \Omega {\bf e}_Z$, where the
direction ${\bf e}_Z$ of the reference frame (${\bf e}_X, {\bf
e}_Y, {\bf e}_Z$) is vertical (see Fig.~\ref{fig:frame}). Fluid
particles forced to oscillate with a pulsation $\sigma< 2\Omega$
describe anticyclonic circular trajectories in tilted planes. A
propagating wave defined by a wavevector ${\bf k}$ normal to these
oscillating planes is solution of the linearized inviscid
equations, satisfying the dispersion relation:
\begin{equation}
\sigma = 2 {\boldsymbol \Omega} \cdot {\bf k} / k = 2 \Omega \cos \theta.
\label{eq:dr}
\end{equation}
In this relation, only the angle of ${\bf k}$ with respect to the
rotation axis is prescribed, whereas its magnitude is set by the
boundary conditions. For such anisotropic dispersion relation, the
phase velocity, ${\bf c} = \sigma {\bf k} / k^2$, is normal to the
group velocity,\cite{Lighthill1978} ${\bf c}_g = \nabla_{\bf k}
\sigma$ (see Fig.~\ref{fig:frame}).

If one now considers a wave forced by a thin horizontal velocity
disturbance invariant in the $Y$ direction, although the velocity
field is still 3 components, the wave pattern is two-dimensional,
varying only in the $(X,Z)$ vertical plane. The wave pattern
consists in four plane beams making angle $\pm \theta$ with
respect to the horizontal, drawing the famous {\it St. Andrew's
cross} familiar in the context of internal
waves.\cite{Mowbray1967} In the following, we consider only one of
those four beams, with $X>0$ and $Z>0$, and we define in
Fig.~\ref{fig:frame} the associated local system of coordinates
(${\bf e}_x, {\bf e}_y, {\bf e}_z$): The axis ${\bf e}_x$ is in
the direction of the group velocity, ${\bf e}_z$ is directed along
the wavevector ${\bf k}$, and ${\bf e}_y = {\bf e}_Y$ is along the
wavemaker.

Since the source is localized, a broad spectrum of wavevectors is
excited, all aligned with ${\bf e}_z$. In an inviscid fluid, the
interference of this infinite set of plane waves will cancel out
everywhere except in the $z=0$ plan, resulting in a single thin
oscillating sheet of fluid describing circular trajectory normal
to ${\bf e}_z$.

\subsection{Viscous spreading}

In a viscous fluid, the energy of the wave beam is dissipated
because of the shearing motion between oscillating planes. As the
energy propagates away from the source, the larger wavenumbers
will be damped first, so that the spectrum of the wave beam
gradually concentrates towards lower wavenumbers, resulting in a
spreading of the wave beam away from the source.

Although the viscous attenuation of a single Fourier component
yields a  purely exponential decay, the attenuation of a localized
wave follows a power law with the distance from the source, which
originates from the combined exponential attenuation of its
Fourier components. A similarity solution for the viscous
spreading of a wave beam has been derived by Thomas and
Stevenson\cite{Thomas1972} in the case of internal waves, and has
been extended to the case of coupled internal-inertial waves by
Peat.\cite{Peat1978} The derivation in the case of a pure inertial
wave is detailed in the Appendix, and we provide here only a
qualitative argument for the broadening of the wave beam.

During a time $t$, the amplitude of a planar monochromatic wave of
wavevector ${\bf k}$ is damped by a factor $\epsilon_k = \exp(-\nu
k^2 t)$ as it travels of a distance $x = c_g t$ along the beam,
where $c_g$ is the group velocity. Using $c_g = (2 \Omega / k)
\sin \theta = (\sigma / k) \tan \theta$, the attenuation factor
writes
$$
\epsilon_k = \exp(- \ell^2 k^3 x),
$$
where we introduce the viscous lengthscale,
\begin{equation}
\ell = \left( \frac{\nu}{\sigma \tan \theta} \right)^{1/2}.
\label{eq:ell}
\end{equation}
For a wave beam emitted from a thin linear source at $x=0$, an
infinite set of plane waves are generated, and the energy of the
largest wavenumbers will be preferentially attenuated as the wave
propagates in the $x$ direction. At a distance $x$ from the
source, the largest wavenumber, for which the energy has decayed
by less than a given factor $\epsilon^*$, is $k_{max} = (\ell^2
x)^{-1/3} \ln \epsilon^*$. At this distance $x$, the wave beam
thus results from the interference of the remaining plane waves of
wavenumbers ranging from 0 to $k_{max}$. Its thickness can be
approximated by $\delta(x) \sim k_{max}^{-1}$, yielding $\delta(x)
/ \ell \sim (x / \ell)^{1/3}$. Mass conservation across a
surface normal to the group velocity implies that the velocity amplitude of
the wave must decrease as $x^{-1/3}$.

More specifically, introducing the reduced transverse coordinate
$\eta = z / x^{1/3} \ell^{2/3}$, a similarity solution exists for
the velocity envelope,
\begin{equation}
u_0(x) = U_0^* \frac{E_0 (\eta)}{E_0 (0)}
\left(\frac{\ell}{x}\right)^{1/3}, \label{eq:u0}
\end{equation}
where $U_0^*$ is the velocity scale of the wave, and the
analytical expression of the non-dimensional envelope $E_0 (\eta)$
is given in the Appendix.  Similarly, the vorticity envelope can be written as
\begin{equation}
\omega_0 (x) = W_0^* \frac{E_1 (\eta)}{E_1(0)}
\left(\frac{\ell}{x}\right)^{2/3}, \label{eq:w0}
\end{equation}
with $W_0^*$ the vorticity scale. Although the normalized velocity
envelope $E_0(\eta) / E_0(0)$ has larger tails than the vorticity
one $E_1(\eta) / E_1(0)$, they turn out to be almost equal for
$\eta < 4$. The width at mid-height, defined such that
$E_m(\eta_{1/2}/2) = E_m (0)/2$, with $m=0,1$, is $\eta_{1/2}
\simeq 6.84$ for both envelopes, so that the width of the beam in
dimensional units is
\begin{equation}
\delta(x) \simeq 6.84 \, \ell \left(\frac{x}{\ell}\right)^{1/3}.
\label{eq:delta}
\end{equation}

\subsection{Finite size effect of the source}
\label{sec:fse}

The similarity solution described here applies only in the case of
a source of size much smaller than the viscous scale $\ell$. In
the case of internal waves, Hurley and Keady\cite{Hurley1997} (see
also Flynn {\it et al.}\cite{Flynn2003}) have shown that for a
source of large extent, vertically vibrated with a small
amplitude, the wave could be approximately described as
originating from two virtual sources, respectively located at the
top and bottom of the disturbance. Following qualitatively this
approach in the case of inertial waves forced by a horizontal
cylinder of radius $R$, the boundaries of the upper wave are given
by $z_\mathrm{up}^\pm = R \pm \delta(x)/2$, and that of the lower
wave are given by $z_\mathrm{down}^\pm = - R \pm \delta(x)/2$. The
lower boundary of the upper source intersects the upper boundary
of the lower source at a distance $x_i$ such that $z_\mathrm{up}^-
(x_i) = z_\mathrm{down}^+ (x_i)$, yielding
\begin{equation}
\frac{x_i}{R} \simeq 0.025\, \left(\frac{R}{\ell}\right)^2.
\label{eq:xstar}
\end{equation}
For large wavemakers ($R / \ell \gg 0.025^{-1/2} \simeq 6.3$), one
has two distinct wave beams for $x \ll x_i$, and one single merged
beam for $x \gg x_i$. On the other hand, for smaller wavemakers,
the merging of the two wave beams occur virtually inside the
source, which can be effectively considered as a point source. In
this case, the effective beam width far from the source may be
simply written as
\begin{equation}
\delta_\mathrm{eff}(x) \simeq 2R + \delta(x).
\label{eq:deff}
\end{equation}

\section{The experiment}
\label{sec:expe}

\subsection{Experimental setup}

\begin{figure}
\psfrag{ex}[c][][1]{${\bf e}_X$} \psfrag{ey}[c][][1]{${\bf e}_Y$}
\psfrag{ez}[c][][1]{${\bf e}_Z$} \psfrag{L}[c][][0.6]{$60$ cm}
\psfrag{H}[c][][0.6]{$54$ cm} \psfrag{O}[c][][0.8]{${\bf \Omega}$}
\psfrag{h1}[c][][0.6]{$34.5$ cm} \psfrag{h2}[c][][0.6]{$h'=33.5$
cm} \psfrag{h3}[c][][0.6]{$8$ mm}
\psfrag{S}[l][][0.9]{\color{red}{$Z_0(t)=A\,\cos(\sigma_o\,t)$}}
\centerline{\includegraphics[width=7.5cm]{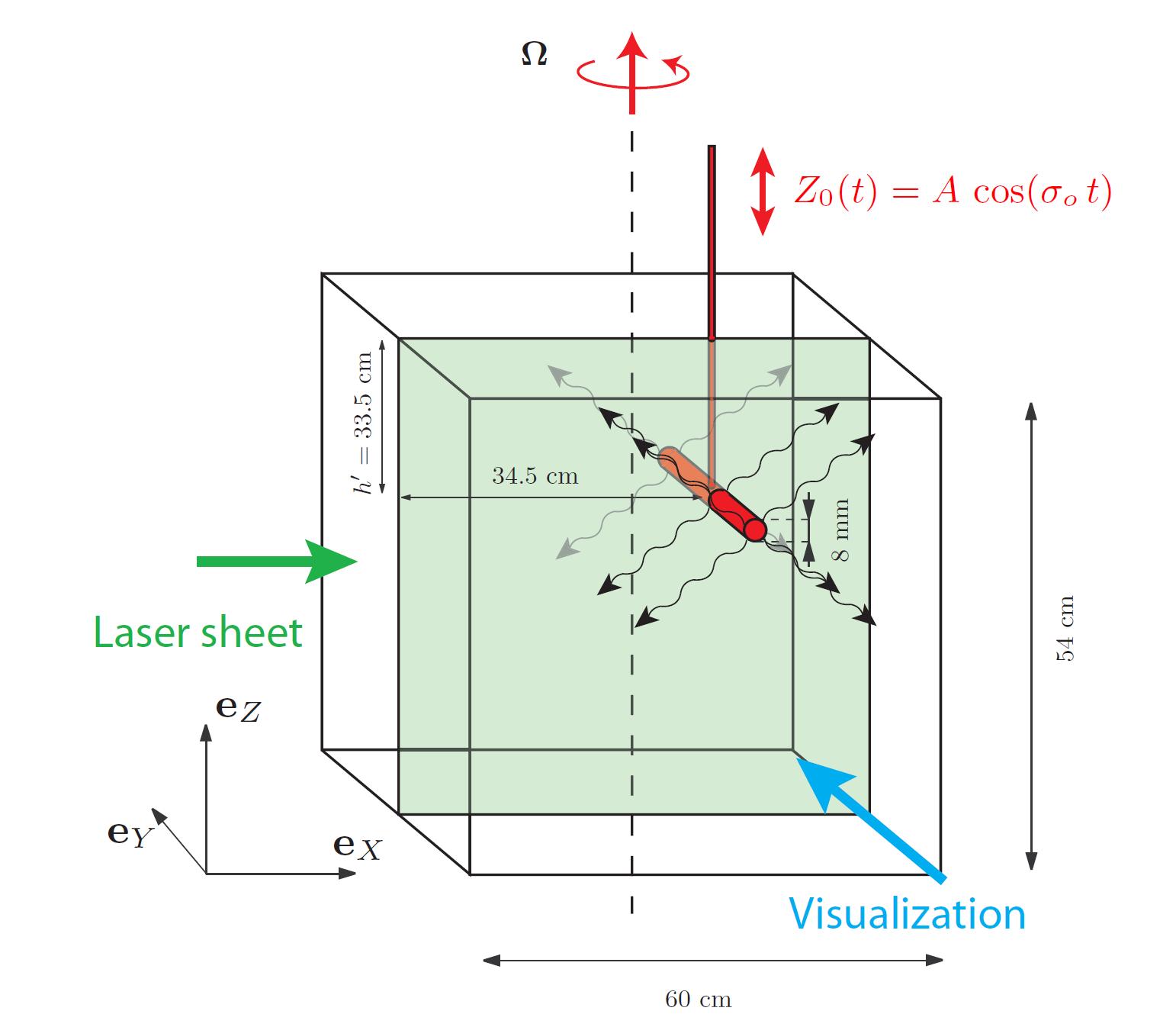}}
\caption{Schematic view of the experimental setup. The horizontal
8 mm diameter cylinder is oscillating vertically according to
$Z_0(t)=A\,\cos(\sigma_o\,t)$, with $A=2$~mm and
$\sigma_o=0.2$~Hz. PIV measurements in a vertical plane ($X$,$Z$)
in the rotating frame are achieved by a vertical laser sheet and a
camera at $90^\circ$.} \label{fig:setup}
\end{figure}

The experimental setup consists in a cubic glass tank, of sides
$60$~cm and filled with 54~cm of water (see Fig.~\ref{fig:setup}),
mounted on the new precision rotating turntable ``Gyroflow'',
of 2~m in diameter.  The
angular velocity $\Omega$ of the turntable is set in the range
0.63 to 2.09~rad~s$^{-1}$, with relative fluctuations $\Delta
\Omega / \Omega$ less than $5 \times 10^{-4}$. A cover is placed
at the free surface, preventing from disturbances due to residual
surface waves. The rotation of the fluid is set long before each
experiment in order to avoid transient spin-up recirculation flows
and to achieve a clean solid body rotation.

The wavemaker is a horizontal cylinder of radius $R=4$~mm and
length $L=50$~cm, hung at $33.5$~cm below the cover by a thin
vertical stem of 3~mm diameter. It is off-centered in order to
increase the size of the studied wave beam in the quadrant $X<0$
and $Z<0$. The vertical oscillation $Z_0(t) = A\, \cos (\sigma_o
t)$, with $A=2$~mm, is achieved by a step-motor, coupled to a
circular camshaft which converts the rotation into a sinusoidal
vertical oscillation. In the present experiments, the wavemaker
frequency is kept constant, equal to $\sigma_o =
1.26$~rad~s$^{-1}$, and the angular velocity of the turntable is
used as the control parameter. This allows the velocity
disturbance $\sigma_o \, A = 2.5$~mm~s$^{-1}$  to be fixed,
whereas the angle of the inertial wave beam with respect to the
horizontal, $\theta = \cos^{-1} (\sigma_o / 2\Omega)$, is varied
between 0 and 72$^\mathrm{o}$. The velocity and vorticity profiles
are examined at distances $x$ between 30 and 300~mm from the
wavemaker. The three-dimensional effects originating from the
finite size $L$ of the cylinder can be safely neglected since
$x < 0.6 L$. The Reynolds number based on the wavemaker velocity is
$Re = \sigma_o A (2R) / \nu \simeq 20$, so that the flow in the
vicinity of the wavemaker is essentially laminar. Except in
Sec.~\ref{sec:te}, where the transient regime is described,
measurements start after several wavemaker periods in order to
achieve a steady state.

For the forcing frequency $\sigma_o$ considered here, the boundary
layer thickness is $\delta_S = (\nu / \sigma_o)^{1/2} \simeq
0.9$~mm. This thickness also gives the order of magnitude of the
viscous length $\ell = \delta_S/\sqrt{\tan \theta}$ [see
Eq.~(\ref{eq:ell})], for angles not too close to 0 and $\pi / 2$.
The wavemaker radius being chosen such that $R/\ell \simeq 4$, the
small source approximation is satisfied according to the criterion
discussed in Sec.~\ref{sec:fse}.

\subsection{PIV measurements}
\label{sec:piv}

Velocity fields in a vertical plane ($X$,$Z$) are measured using a
2D Particle Image Velocimetry system. The flow is seeded by
10~$\mu$m tracer particles, and illuminated by a vertical laser
sheet, generated by a 140~mJ Nd:YAg pulsed laser. A vertical
$43\times43$~cm$^2$ field of view is acquired by a $2048 \times
2048$ pixels camera synchronized with the laser pulses. The field
of view is set on the lower left wave beam. For each rotating
rate, a set of 2000 images is recorded, at a frequency of 2 Hz,
representing 10 images per wavemaker oscillation period.

PIV computations are performed over successive images, on
$32\times32$ pixels interrogation windows with $50\%$ overlap,
leading to a spatial resolution of 3.4~mm.\cite{DaVis}
In the following, the
two quantities of interest are the velocity component $u_x$,
obtained from the measured components $u_X$ and $u_Z$ projected
along the direction of the wave beam, and the vorticity component
$\omega_y$, normal to the measurement plane.

The velocity along the wave beam typically decreases from 1 to
0.1~mm~s$^{-1}$, and is measured with a resolution of
0.02~mm~s$^{-1}$. Two sources of velocity noise are present, both
of order of 0.2~mm~s$^{-1}$, originating from residual modulations
of the angular velocity of the turntable, and from thermal
convection effects due to a slight difference between the water
and the room temperature. The residual velocity modulations, of
order of $L_0 \, \Delta \Omega / 2$ (where $L_0$ is the tank size and
$\Delta \Omega \simeq 5 \times 10^{-4} \Omega$) are readily
removed by computing the phase-averaged velocity fields over the
200 periods of the wavemaker.  Thermal convective motions, in the
form of slowly drifting ascending and descending columns, could
be reduced but not completely suppressed  by this phase-averaging,
and represent the main source of uncertainty in these experiments.
However, the vorticity level associated to those
convective motions appears to be negligible compared to the
typical vorticity of the inertial wave. Therefore, the vorticity
profiles of the wave could be safely computed from the
phase-averaged velocity fields.

\section{General properties of the wave pattern}

\subsection{Visualisation of the wave beams}

\begin{figure}
\psfrag{X}[c][][0.8]{$X$ (mm)}\psfrag{Y}[c][][0.8]{$Z$ (mm)}
\psfrag{W}[l][][0.8]{$\omega$
(s$^{-1}$)}\psfrag{P1}[l][][0.8]{$\Phi=0$}\psfrag{P2}[l][][0.8]{$\phi=2\pi/5$}
\psfrag{P3}[l][][0.8]{$\phi=4\pi/5$}\psfrag{P4}[l][][0.8]{$\phi=6\pi/5$}
\begin{center}
\centerline{\includegraphics[width=8cm]{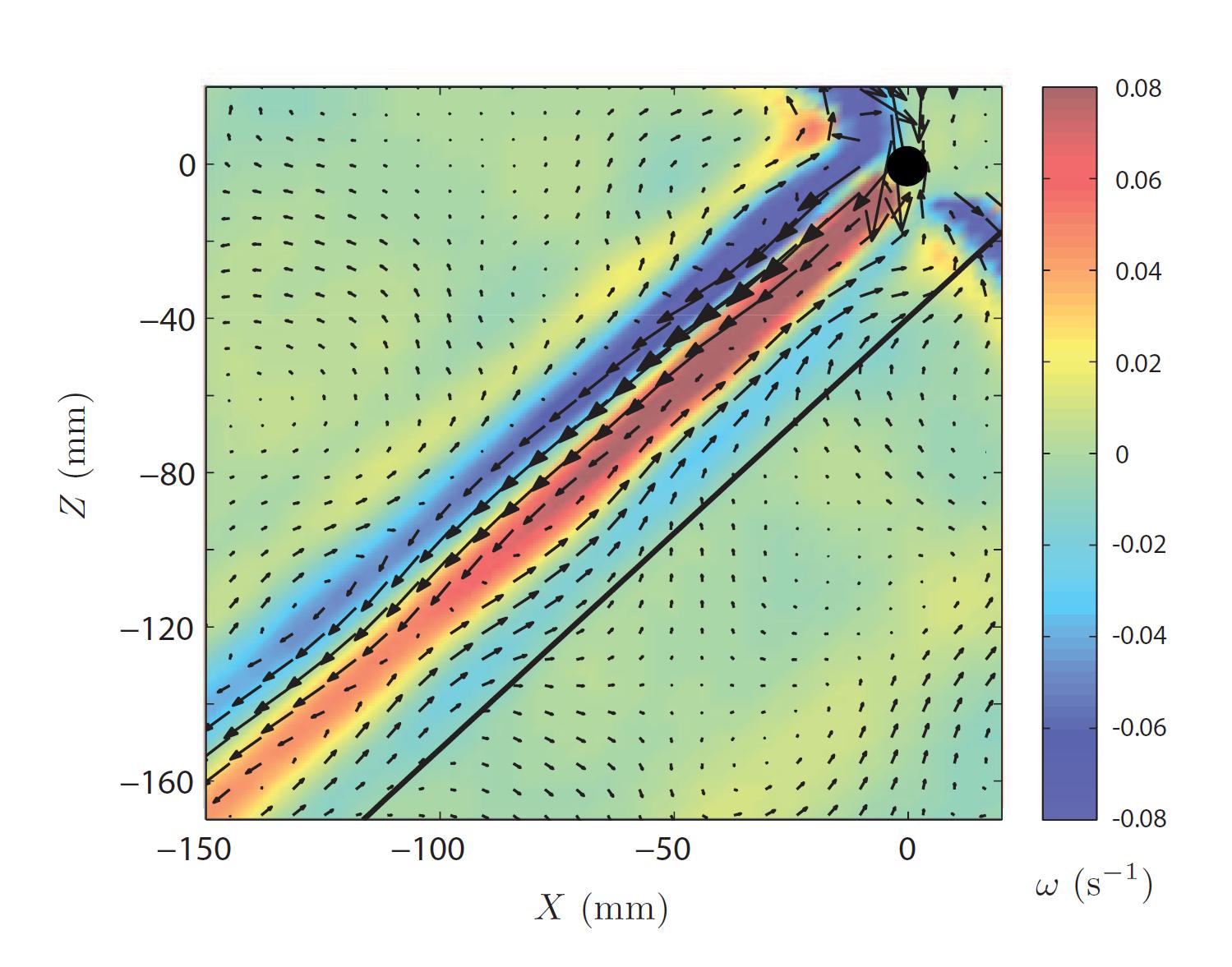}}
\caption{Close-up view of the phase averaged velocity (arrows) and
vorticity $\omega_y$ (color mapped) for an experiment performed at
$\sigma_o/2\Omega=0.67$. The black line shows the direction
predicted by the dispersion relation $\cos
\theta=\sigma_o/2\Omega$.} \label{fig:champvit}
\end{center}
\end{figure}

\begin{figure}
\psfrag{X}[c][][0.8]{$X$ (mm)}\psfrag{Y}[c][][0.8]{$Z$ (mm)}
\psfrag{C}[c][][0.8]{${\bf c}$} \psfrag{W}[c][][0.7]{$\omega$
(rad\,s$^{-1}$)}\psfrag{P1}[l][][0.8]{(a)\hspace{0.2cm}$\phi=\pi/5$}\psfrag{P2}[l][][0.8]{(b)\hspace{0.2cm}$\phi=2\pi/5$}
\psfrag{P3}[l][][0.8]{(c)\hspace{0.2cm}$\phi=3\pi/5$}\psfrag{P4}[l][][0.7]{(d)
Vorticity envelope}
\begin{center}
\centerline{\includegraphics[width=8.5cm]{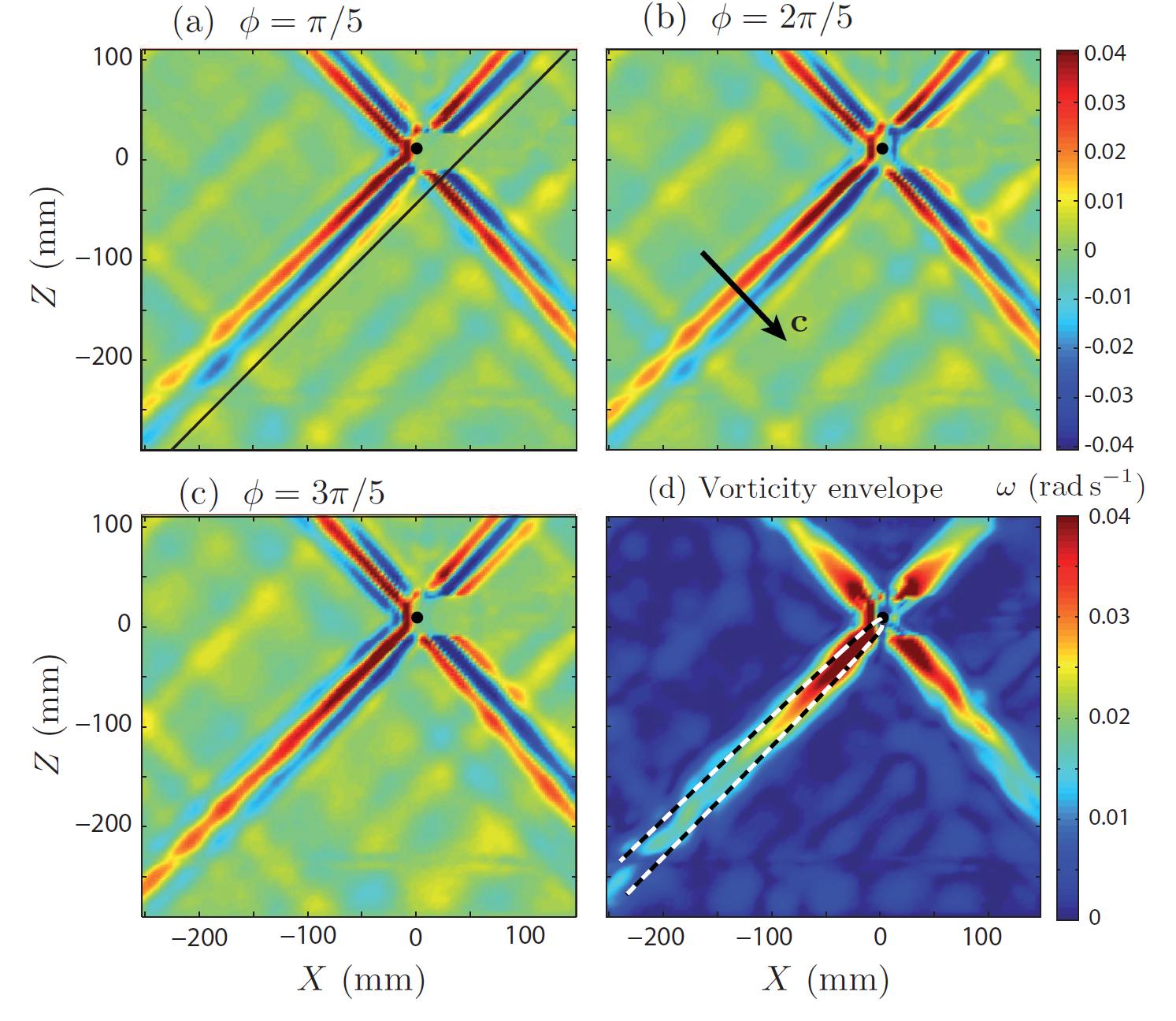}}
\caption{Phase averaged horizontal vorticity field $\omega_y$ for
$\sigma_o/2\Omega=0.67$ at different phases: (a) $\phi=\pi/5$, (b)
$\phi=2\pi/5$ and (c) $\phi=3\pi/5$. The black line in (a) draws
the direction predicted by the dispersion relation. (d) Vorticity
envelope field $\omega_0$ (see Sec.~\ref{sec:vp}). The dashed
black and white lines shows the wave beam thickness predicted by
the similarity solution [see Eq. (\ref{eq:deff})].}
\label{fig:champvort}
\end{center}
\end{figure}

Figure \ref{fig:champvit} is a close-up view of the velocity and
vorticity fields at $\sigma_o/2\Omega=0.67$, showing the vorticity
layers of alternating sign where the measured velocity is almost
parallel to the ray direction ${\bf e}_x$. The angle of the ray
with respect to the horizontal accurately follows the prediction
of the dispersion relation (\ref{eq:dr}), as illustrated by the
black line. In Fig.~\ref{fig:champvort}(a) to (c),  phase-averaged
horizontal vorticity fields $\omega_y$ are shown for three
regularly sampled values of the phase. One can clearly see the
location of the inertial wave inside a wavepacket that draws the
classical four-rays \textit{St. Andrew's cross}.  The evolution of
the vorticity field from Fig.~\ref{fig:champvort}(a) to (c)
illustrates the propagation of the phase in directions normal to
the rays. Some reflected wave beams of much smaller amplitude may
also be distinguished on the background.

\begin{figure}
\psfrag{X}[c][][1]{$\sigma_o/2\Omega$} \psfrag{Y}[c][][1]{$\cos
\theta$}
\begin{center}
\centerline{\includegraphics[width=7cm]{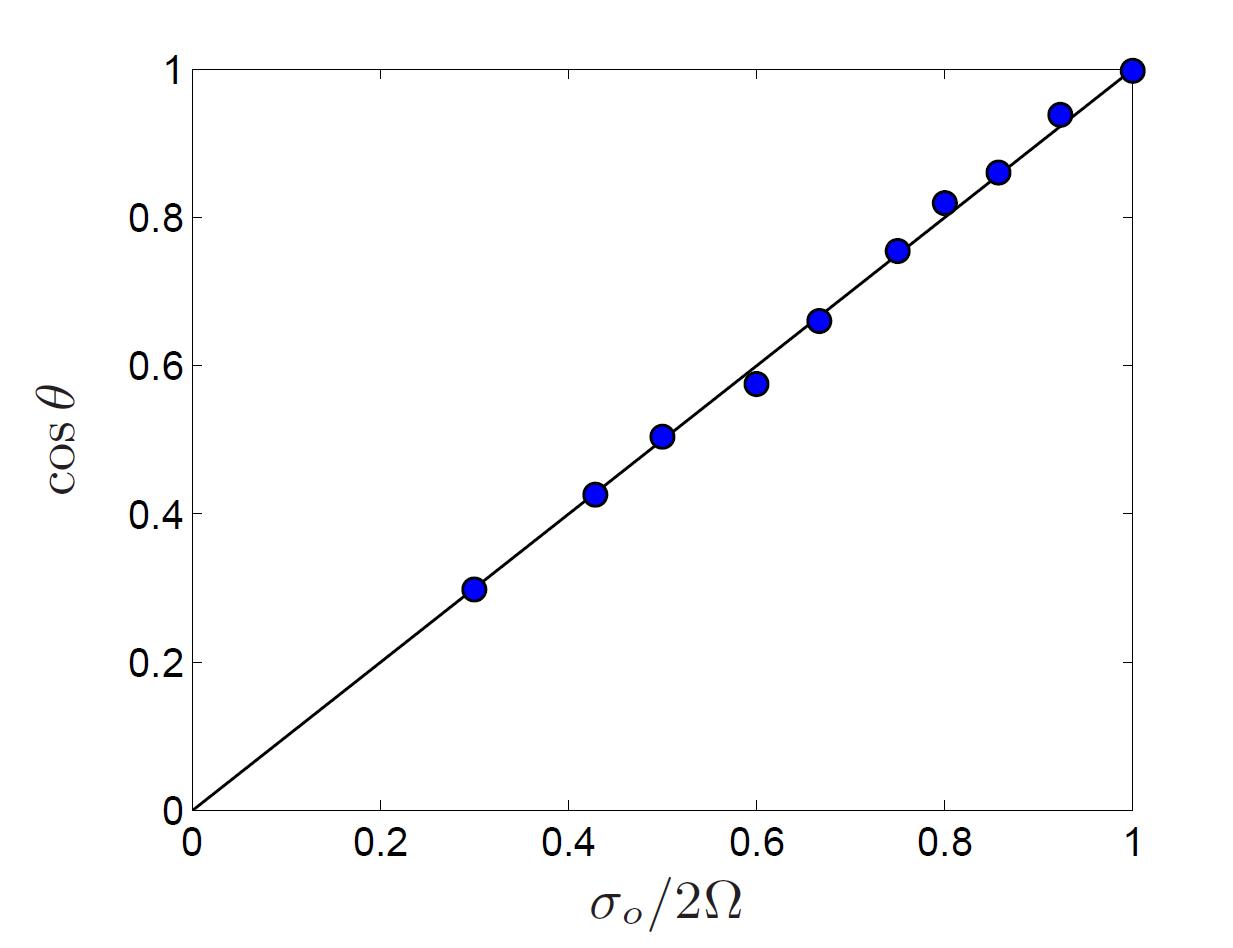}}
\caption{Cosine of the measured average beam angle, $\cos \theta$,
as a function of the frequency ratio $\sigma_o/2\Omega$. The line
shows the dispersion relation (\ref{eq:dr}). Experimental
uncertainties are of the order of the marker size.}
\label{fig:disp}
\end{center}
\end{figure}

Figure~\ref{fig:disp} compares the cosine of the measured angle
$\theta$ with the theoretical value $\sigma_o/2\Omega$ according
to the prediction of the dispersion relation (\ref{eq:dr}). The
angle is determined from the location of the maximum of the
vorticity envelope (envelope computations described in
Sec.~\ref{sec:vp}), shown in Fig.~\ref{fig:champvort}(d), averaged
along the ray. An excellent agreement is obtained between the
measurements and the theory, with a relative error less than
$3\%$.

\subsection{Transient experiments}
\label{sec:te}

In order to characterize the formation of the inertial wave
pattern as the oscillation is started, a series of transient
experiments has been performed. In the case of a pure
monochromatic plane wave, the front velocity of the wavepacket is
simply given by the group velocity. However, in the case of a
localized wave beam, since each Fourier component $k$ travels with
its own group velocity $c_g = (\sigma / k) \tan \theta$, the shape
of the wavepacket gradually evolves as the wave propagates. A
rough estimate for the front velocity can be simply obtained from
$V_f \simeq \sigma (\lambda/2\pi) \tan \theta$, where
$\lambda$ is the apparent wavelength of the wave, simply estimated
as twice the distance between the locations of two successive
vorticity extrema.

\begin{figure}
\psfrag{Y}[c][][0.8]{$time$ (s)}\psfrag{X}[c][][1]{$x$
(mm)}\psfrag{W}[l][][0.8]{$\omega$ (m.s$^{-1}$)}
\psfrag{SO1}[l][][0.8]{(a)\hspace{0.2cm}$\sigma_o/2\Omega=0.85$}\psfrag{SO2}[l][][0.8]{(b)\hspace{0.2cm}$\sigma_o/2\Omega=0.80$}
\psfrag{SO3}[l][][0.8]{(c)\hspace{0.2cm}$\sigma_o/2\Omega=0.75$}\psfrag{SO4}[l][][0.8]{(d)\hspace{0.2cm}$\sigma_o/2\Omega=0.67$}
\psfrag{SO5}[l][][0.8]{(e)\hspace{0.2cm}$\sigma_o/2\Omega=0.60$}\psfrag{SO6}[l][][0.8]{(f)\hspace{0.2cm}$\sigma_o/2\Omega=0.50$}
\begin{center}
\centerline{\includegraphics[width=8.5cm]{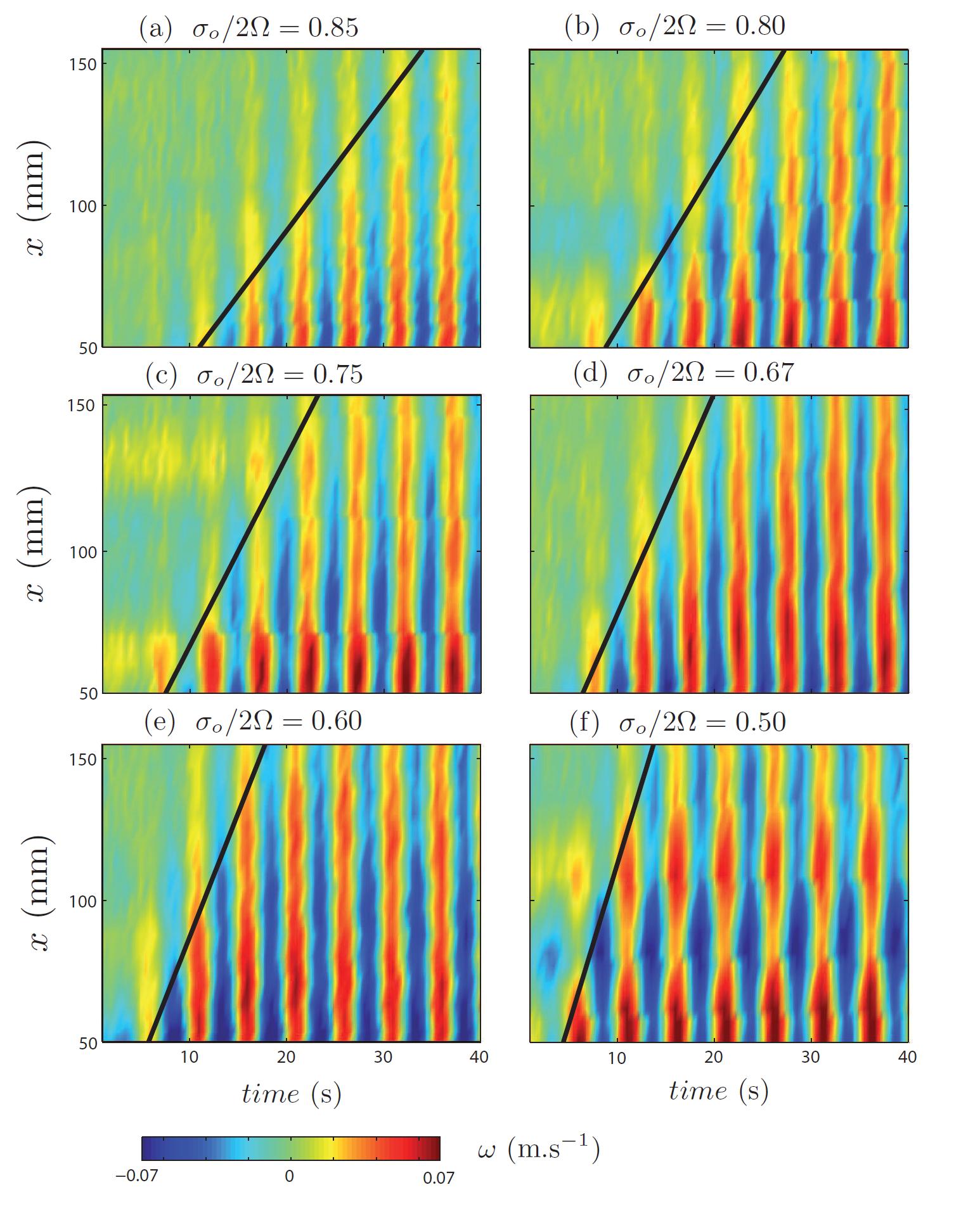}}
\caption{Spatiotemporal representation of the vorticity $\omega_y$
along the wave beam, where space is the distance $x$ to the
oscillating cylinder, for experiments performed at
$\sigma_o/2\Omega=0.85,\,0.80,\,0.75,\,0.66,\,0.60,\,0.50$. Black
lines originating at ($x=0,t=0$) trace the front velocity
$V_f=\sigma (\lambda/2\pi) \tan \theta$ estimated from the
apparent wavelength (see Sec. \ref{sec:vve}). $t=0$ corresponds to
the start of the oscillation.} \label{fig:vgroup}
\end{center}
\end{figure}

Figure~\ref{fig:vgroup} shows spatiotemporal diagrams of the
vorticity $\omega_y(x,z=0,t)$ at the center of the beam  as a
function of the distance $x$ from the wavemaker,  for
$\sigma_o/2\Omega$ between 0.85 and 0.50. Superimposed to these
spatiotemporal images, we show the front velocity $V_f \simeq
\sigma ( \lambda/2\pi) \tan \theta$, starting from $x=0$ at
$t=0$. Qualitative agreement with the spatiotemporal diagrams is
obtained, indicating that the propagation of the wave envelope is
indeed compatible with this front velocity.

Further quantitative estimate of the front velocity would require
to extract the instantaneous wave envelope from those
spatio-temporal diagrams, which is difficult because the front
velocity and the phase velocity are of the same order. This
property actually prevents from a safe extraction of a
longitudinal wavepacket envelope using standard temporal averaging
over small time windows.

\subsection{Generation of harmonics}

Going back to steady waves, we now characterize the generation of
higher order wave beams which take place at low forcing frequency.
According to the dispersion relation, an harmonic wave of order
$n \geq 2$ is allowed to develop whenever $n \sigma_o/ 2 \Omega<1$. Such
harmonic waves of order $n \geq 2$ may originate either from a residual
non-harmonic component of wavemaker oscillation profile $Z_0(t)$,
or from inertial non-linear effects in the flow in the vicinity of
the wavemaker, which can exist at the Reynolds number $Re \simeq
20$ considered here.

\begin{figure}
\psfrag{a}[l][][1]{(a)} \psfrag{b}[l][][1]{(b)}
\psfrag{c}[l][][1]{(c)} \psfrag{X}[c][][0.8]{$X$
(mm)}\psfrag{Y}[c][][0.8]{$Z$ (mm)} \psfrag{W}[c][][0.8]{$\omega$
(s$^{-1}$)}\psfrag{P1}[l][][0.8]{$\Phi=0$}\psfrag{P2}[l][][0.8]{$\Phi=2\pi/5$}
\psfrag{P3}[l][][0.8]{$\Phi=4\pi/5$}\psfrag{P4}[l][][0.8]{$\Phi=6\pi/5$}
\begin{center}
\centerline{\includegraphics[width=9cm]{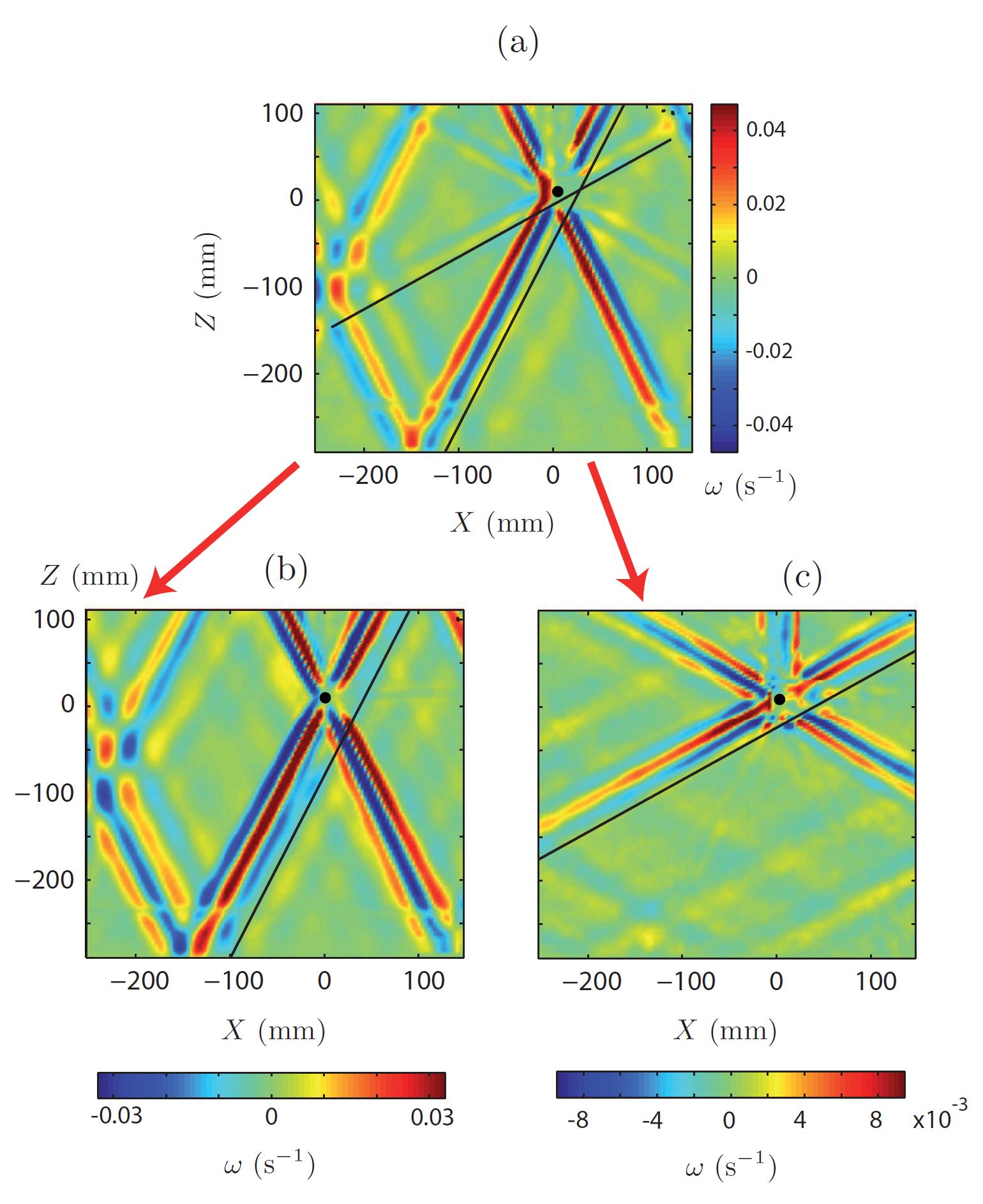}}
\caption{(a) Phase averaged vorticity field $\omega_y$ for an
experiment performed at $\sigma_o/2\Omega=0.43$, showing both the
fundamental ($n=1$) and the second harmonic ($n=2$) wave beams.
The corresponding frequency-filtered vorticity fields are
extracted in (b) and (c).} \label{fig:champbis}
\end{center}
\end{figure}

In Fig.~\ref{fig:champvort}, in which $\sigma_o/ 2 \Omega=0.67$,
only the fundamental wave ($n=1$) can be seen. On the other hand,
in Fig.~\ref{fig:champbis}(a), in which $\sigma_o/ 2 \Omega=0.43$,
a second harmonic wave beam is clearly present, propagating at an
angle closer to the horizontal,  as expected from the dispersion
relation. This is confirmed by Fig.~\ref{fig:champbis}(b) and (c),
showing the corresponding frequency-filtered phase averaged
vorticity fields, in (b) for the fundamental $n=1$ and in (c) for
the second harmonics $n=2$.

\begin{figure}
\psfrag{a}[l][][1]{(a)} \psfrag{b}[l][][1]{(b)}
\psfrag{n1}[l][][0.9]{{\color{red}$n=1$}}
\psfrag{n2}[l][][0.9]{{\color{blue}$n=2$}}
\psfrag{X}[c][][1]{$\sigma/2\Omega$} \psfrag{Y}[c][][1]{$|\hat
u_\sigma|^2$ (m$^2$~s$^{-1}$)}
\psfrag{FS}[c][][0.9]{$\sigma=\sigma_o$}
\psfrag{2FS}[l][][0.9]{$\sigma=2\sigma_o$}
\psfrag{2O}[c][][0.9]{$\sigma=\Omega$}
\psfrag{4O}[c][][0.9]{$\sigma=2\Omega$}
\begin{center}
\centerline{\includegraphics[width=7.5cm]{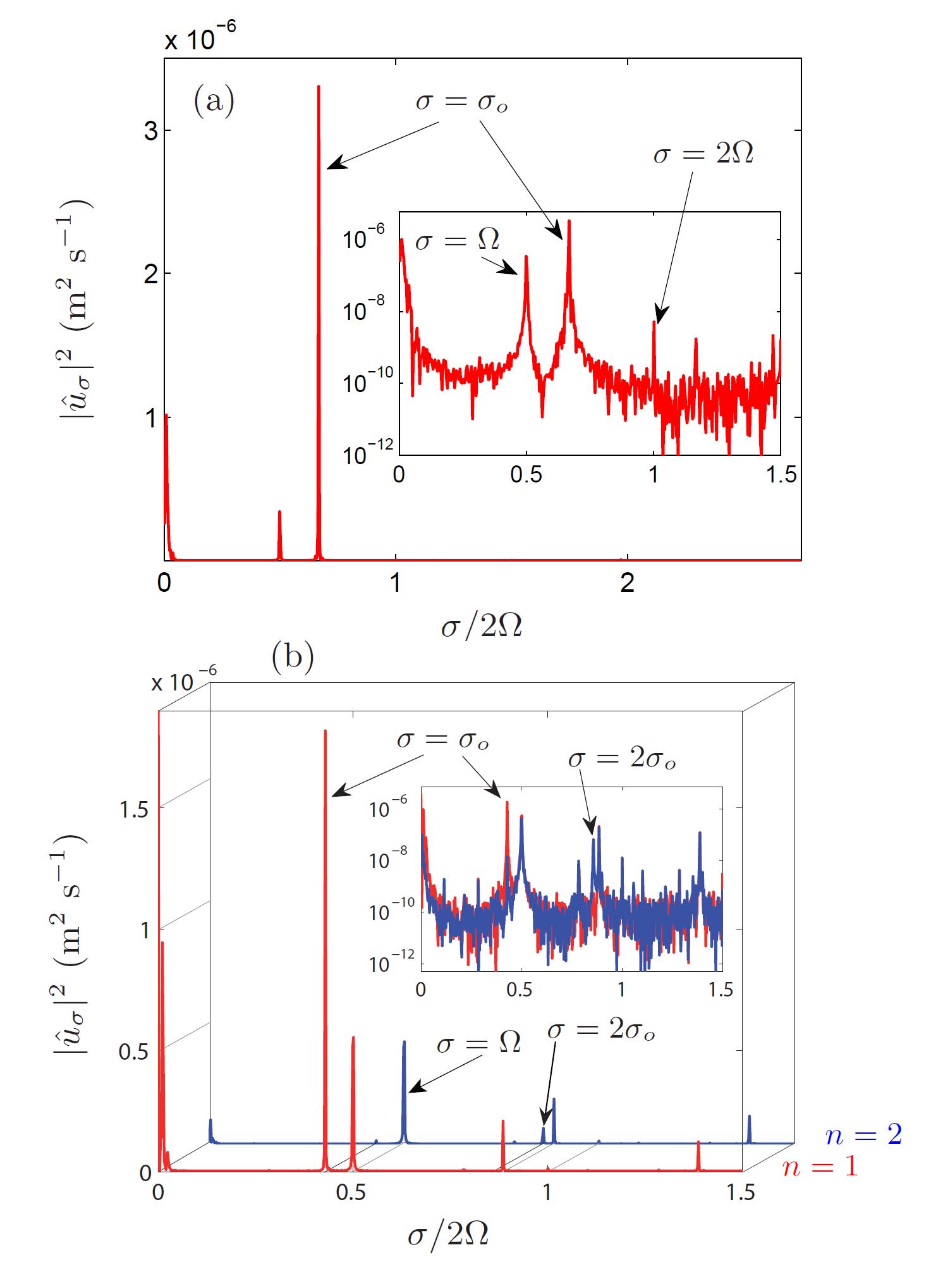}}
\caption{Energy spectrum of the velocity time series measured at
the center of the wave beam of interest, at a fixed distance $x_0
= 100$~mm from the wavemaker. (a) $\sigma_o/2\Omega=0.67$, showing
a single peak at the forcing frequency. (b)
$\sigma_o/2\Omega=0.43$, showing measurements performed in the
fundamental beam $n=1$ (in red/light gray) and in the second
harmonic beam $n=2$ (in blue/dark gray). In (a) and (b), insert
shows the same spectrum in semi-logarithmic coordinates.
Additional peaks are present at $\sigma/2\Omega=0.5$ and $1$,
originating from mechanical noise of the rotating platform.}
\label{fig:spec}
\end{center}
\end{figure}

In order to further characterize this generation of harmonics, we
have performed a spectral analysis of the time series of the
longitudinal velocity $u_x(t)$, measured at a given distance
$x_0=100$~mm from the source, at the center of each wave beam. The
energy spectrum $|\hat u_\sigma |^2$, where $\hat u_\sigma$ is the
temporal Fourier transform of $u_x(t)$, is shown in
Fig.~\ref{fig:spec} for the two cases $\sigma_o / 2 \Omega = 0.67$
and $0.43$. In both cases, the spectra are clearly dominated by
the fundamental forcing frequency $\sigma_o$. Two other peaks are
also found, at $\sigma=\Omega$ and $\sigma=2\Omega$, originating
from the residual modulation of the angular velocity of the
platform, as discussed in Sec.~\ref{sec:piv} (the energy of those
peaks is typically 3 to 10 times smaller than the fundamental
one). As expected, no harmonic frequency $n \sigma_o$ ($n \geq 2$)
is found for $\sigma_o / 2 \Omega = 0.67$ (see
Fig.~\ref{fig:spec}a), but a second harmonic $n=2$ is indeed
present for $\sigma_o/ 2 \Omega=0.43$ (see Fig.~\ref{fig:spec}b).
In this case, the energy ratio of the first to the second
harmonics, each of them being measured at a distance $x_0 =
100$~mm from the source on the corresponding beam, is $|\hat
u_{2\sigma} |^2 / |\hat u_\sigma |^2 \simeq 0.036$. As
$\sigma_o/2\Omega$ is further decreased, the ratio $|\hat
u_{2\sigma} |^2 / |\hat u_\sigma |^2$ increases,  reaching $0.05$
for $\sigma_o/2\Omega=0.30$, and even higher order harmonics
emerge although with very weak amplitude.

\section{Test of the similarity solution}
\label{sec:vp}

\subsection{Velocity and vorticity envelopes}
\label{sec:vve}

\begin{figure}
\psfrag{a}[c][][1]{(a)} \psfrag{b}[c][][1]{(b)}
\psfrag{X}[c][][1]{$z$ (mm)} \psfrag{Y}[c][][1]{$u_x$ (m
s$^{-1}$)} \psfrag{L}[c][][0.9]{$\delta$}
\psfrag{wm}[c][][0.9]{$u_{max}$}
\psfrag{EQ1}[l][][0.85]{}
\psfrag{W}[c][][1]{$\omega_y$ (rad s$^{-1}$)}
\psfrag{wm}[c][][0.9]{}
\psfrag{EQ2}[l][][0.85]{}
\begin{center}
\centerline{\includegraphics[width=8cm]{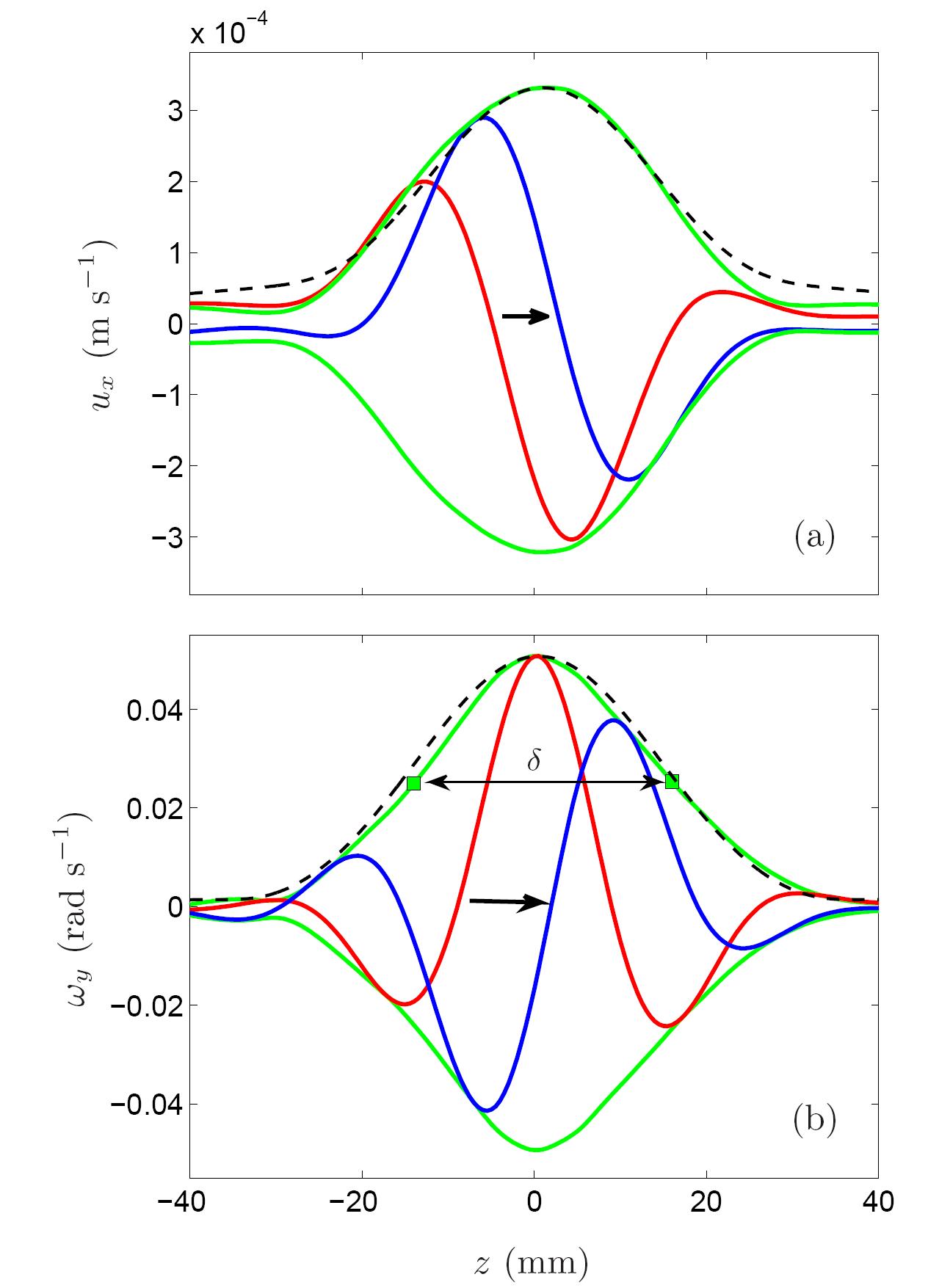}}
\caption{(a) Velocity envelope $u_0 (x_0,z)$ and two velocity
profiles $u_x(x_o,z,\phi)$ for two values of the phase $\phi$, as
a function of the transverse coordinate $z$ at a fixed distance
$x_o=100$~mm from the wavemaker for $\sigma_o/2\Omega=0.67$. (b)
Corresponding vorticity envelope $\omega_0 (x_0,z)$ and vorticity
profiles $\omega_y(x_0,z,\phi)$. $\delta$ is the envelope
thickness at mid-height. In (a) and (b), the green/light gray
curves correspond to the experimental envelope, and the dotted
curves correspond to the predictions of the similarity solutions
normalized by the measured maximum.} \label{fig:env}
\end{center}
\end{figure}

We now focus on the dependence of the wavepacket shape and the
viscous spreading of the wave beam with the distance $x$ from the
source. Figures~\ref{fig:env}(a) and (b) illustrate the shape of
the phase-averaged velocity and vorticity profiles respectively,
for two values of the phase $\phi_0$ and $\phi_0 + 2\pi/5$. The
wavepacket envelopes are defined as
$$
u_{0}(x,z) = \sqrt{2 \langle u_x(x,z,\phi)^2 \rangle_\phi}
$$
(and similarly for $\omega_{0}$), where $\langle \cdot
\rangle_\phi$ is the average over all phases $\phi$. Although the
measured normalized envelopes compare well with the normalized
envelopes predicted from the similarity solutions [$E_m(\eta) /
E_m(0)$, with $m=0$ for the velocity and $m=1$ for the vorticity],
the agreement is actually better for the vorticity. This is probably due to
the velocity contamination originating from the residual angular
velocity modulation of the platform and the slight thermal
convection effects discussed in Sec.~\ref{sec:piv}. The better
defined vorticity envelopes actually confirms that those velocity
contaminations have a negligible vorticity contribution. For this
reason, we will concentrate only on the vorticity field in the
following.

\begin{figure}
\psfrag{X}[c][][1]{$z$ (mm)} \psfrag{V}[c][][1]{$u_x$
(m~s$^{-1}$)} \psfrag{W}[c][][1]{$\omega_y$ (rad~s$^{-1}$)}
\psfrag{a}[c][][1]{(a)} \psfrag{b}[c][][1]{(b)}
\begin{center}
\centerline{\includegraphics[width=7.5cm]{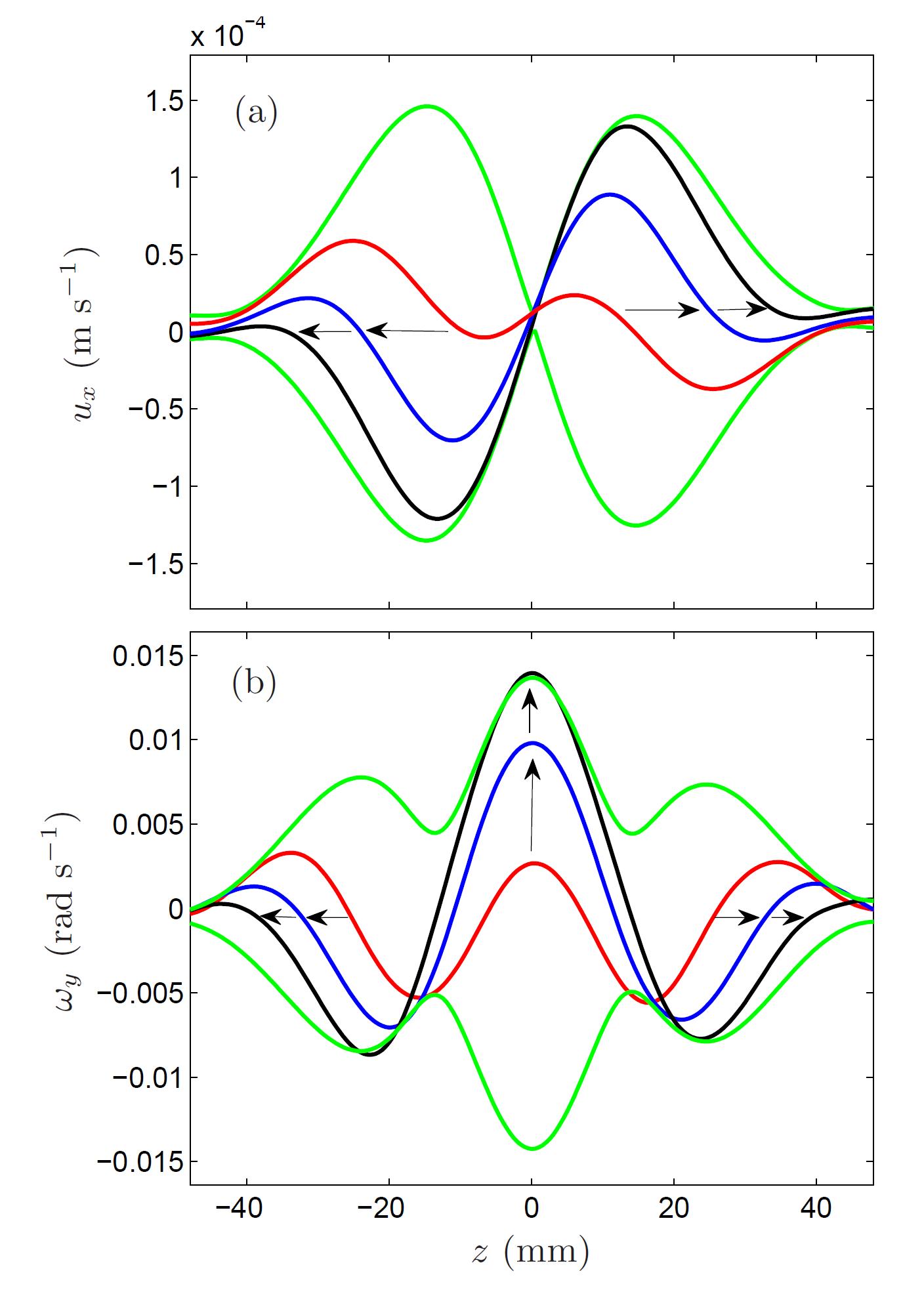}}
\caption{(a) Velocity envelope $u_{0}(x_0,z)$ and three velocity
profiles $u_x(x_0,z,\phi)$ as a function of the transverse
coordinate $z$ at a fixed distance $x_o=70$~mm from the wavemaker
for $\sigma_o/2\Omega=1$. (b) Corresponding vorticity envelope and
vorticity profiles. The arrows indicate the time evolution of the
profiles. The interference of the upward and downward wave beams
produces a stationary wave pattern at $z=0$ with a velocity node
and a vorticity maximum.} \label{fig:envelope2}
\end{center}
\end{figure}

It is worth to examine here the singular situation
$\sigma_o/2\Omega = 1$, in which the similarity solution is no
longer valid. In this situation, the phase velocity is strictly
vertical and the group velocity vanishes. The upward and downward
beams are expected to superimpose and generate a stationary wave
pattern in the horizontal plane $Z=z=0$.
Figure~\ref{fig:envelope2} shows the velocity envelope
$u_{0}(x_0,z)$ and three phase averaged profiles as a function of
the transverse coordinate $z$. The observed wave is actually
stationary at the center of the wavepacket (see the velocity node
and vorticity maximum for $z=0$), and shows outwards propagations
on each side of the wavepacket.

Returning to the standard situation $\sigma_o / 2\Omega < 1$, the
vorticity amplitude at a given location $x$ is defined as the
maximum of the vorticity envelope at the center of the beam,
$\omega_{max}(x) = \omega_{0}(x,z=0)$. The thickness of the
wavepacket $\delta(x)$ is defined from the width at mid-height of
the envelope, such that
$$\omega_{0}(x, \delta(x)/2) = \omega_{max}(x)/2.$$

\begin{figure}
\psfrag{a}[l][][1]{(a)} \psfrag{b}[l][][1]{(b)}
\psfrag{X}[c][][1]{$x$ (mm)} \psfrag{Y}[l][][1]{$\lambda$ (mm)}
\psfrag{S}[c][][1]{$\sigma_o/2\Omega$} \psfrag{Z}[l][][1]{$\delta$
(mm)} \psfrag{EQ}[l][][1]{$\delta_{\rm eff}$} \psfrag{EQ2}[l][][1]{$\delta_{\rm eff}$}
\begin{center}
\centerline{\includegraphics[width=7cm]{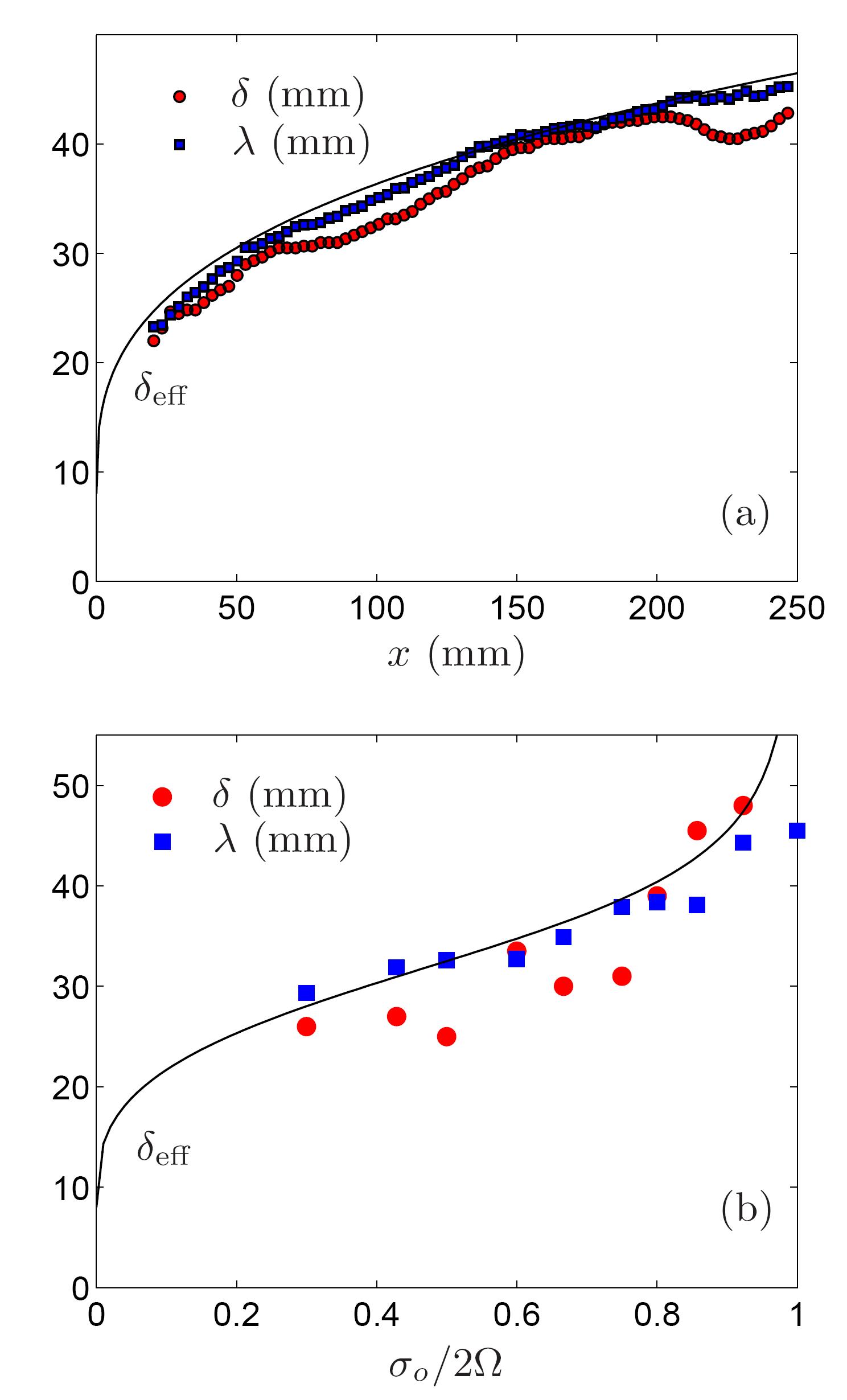}}
\caption{Wave beam thickness $\delta$ (red $\circ$), and apparent
wavelength $\lambda$ (blue $\square$); (a), as a function of the
distance $x$ from the wavemaker for $\sigma_o/2\Omega=0.67$; (b),
as a function of $\sigma_o / 2\Omega$ at a distance $x_o=100$~mm
from the wavemaker. In both plots, the line shows the predicted
effective wave beam thickness $\delta_\mathrm{eff}$
(\ref{eq:deff}).} \label{fig:largeur}
\end{center}
\end{figure}

This beam thickness $\delta$ depends both on the distance $x$ from
the source and on the viscous length $\ell$ [see Eqs.
(\ref{eq:delta}) and (\ref{eq:deff})]. In order to check those two
dependencies, $\delta$ is plotted in Fig.~\ref{fig:largeur}(a) as
a function of $x$ at fixed $\sigma_o / 2\Omega$, and in
Fig.~\ref{fig:largeur}(b) as a function of $\sigma_o / 2\Omega$ at
fixed $x_0$. The agreement with the effective wave beam thickness
$\delta_{\rm eff} = 2R + 6.84 \ell (x/\ell)^{1/3}$ is correct,
to within $10\%$, which justifies
the simple analysis of merged beams originating from the two
virtual sources located at the top and bottom of the wavemaker.
The oscillations of $\delta$ probably originate from the
interaction of the principal wave beam with reflected ones.
Figure~\ref{fig:largeur}(a) also shows the apparent wavelength
$\lambda(x)$ of the wave, simply defined as twice the distance
between a maximum and a minimum of the phase-averaged vorticity
profiles. This apparent wavelength turns out to be even closer to
the expected lengthscale $\delta_{\rm eff}$ of Eq.
(\ref{eq:deff}), to within $4\%$, suggesting that $\lambda$ is
less affected by the background noise than the beam thickness.
A good agreement between both $\delta$ and $\lambda$ and the
prediction (\ref{eq:deff}) is also obtained as $\sigma_o/2\Omega$
(and hence $\ell$) is varied at fixed $x_0$, as shown in
Fig.~\ref{fig:largeur}(b).  Here again, the interaction with
reflected wave beams is probably responsible for the
significant scatter in this figure.

\subsection{Decay of the vorticity envelope}

\begin{figure}
\psfrag{a}[c][][1]{(a)} \psfrag{b}[c][][1]{(b)}
\psfrag{X}[c][][1]{$x$ (mm)}
\psfrag{S}[c][][1]{$\sigma_o/2\Omega$}
\psfrag{W}[c][][1]{$\omega_{max}$ (rad~s$^{-1}$)}
\psfrag{H}[c][][1]{$g(\theta)$} \psfrag{EQ1}[r][][0.8]{$W_0^*
(x/\ell)^{-2/3}$} \psfrag{EQ2}[r][][0.8]{}
\begin{center}
\centerline{\includegraphics[width=7cm]{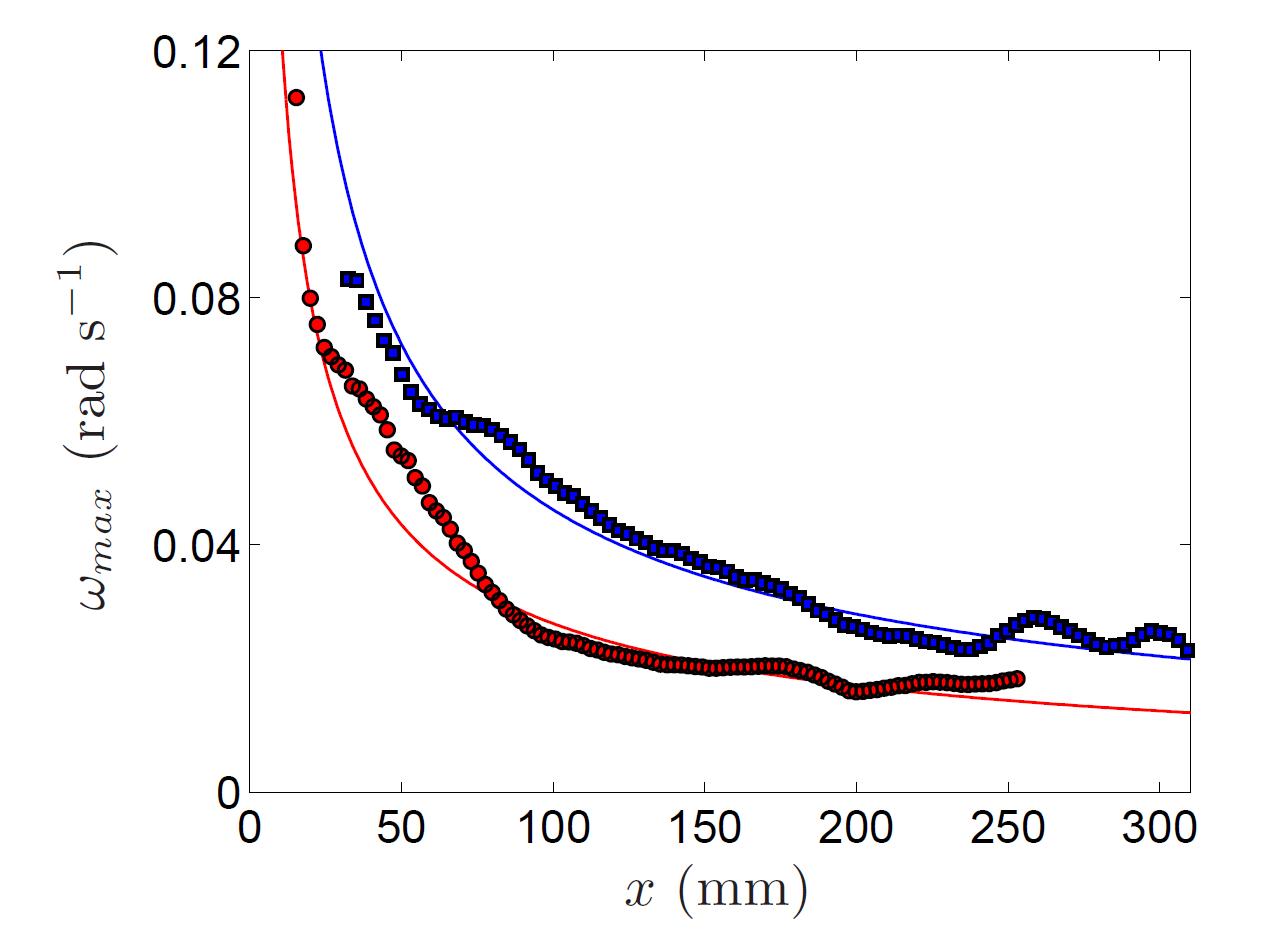}}
\caption{Vorticity amplitude $\omega_{max}(x)$ as a function of
the distance $x$ from the wavemaker for $\sigma_o/2\Omega=0.67$
(blue $\square$) and $\sigma_o/2\Omega=0.85$ (red $\circ$), and
best fit with the law $W_0^* (x / \ell)^{-2/3}$.}
\label{fig:vortampa}
\end{center}
\end{figure}

The decay of the vorticity amplitude $\omega_{max} (x)$ as a
function of the distance $x$ from the source is shown in
Fig.~\ref{fig:vortampa}. Taking the similarity solution
(\ref{eq:w0}) at the center of the wave beam $z = 0$ yields
\begin{equation}
\omega_{max} (x) = W_0^* \, \left(\frac{\ell}{x}\right)^{2/3}.
\label{eq:wmax0}
\end{equation}
Letting the vorticity scale $W_0^*$ as a free parameter, a power
law $x^{-2/3}$ is found to provide a good fit for the overall
decay of $\omega_{max} (x)$. Some marked oscillations are however
clearly visible, e.g. at $x$ between 220 and 320~mm for
$\sigma_o/2\Omega=0.85$ (blue $\square$ in
Fig.~\ref{fig:vortampa}). Those oscillations appear at locations
where reflected wave beams interact with the principal one,
inducing modulations of the wave amplitude. This interpretation is
confirmed by the fact that (i) the observed modulation has a
wavelength of $45$~mm, which corresponds to the apparent
wavelength of the wave, and that (ii) in Fig.~\ref{fig:champvort},
corresponding to $\sigma_o/2\Omega=0.67$, a modulation of the
principal wave beam by a reflected can be clearly seen at a
distance of about $250$~mm from the source.

\begin{figure}
\psfrag{a}[c][][1]{(a)} \psfrag{b}[c][][1]{(b)}
\psfrag{X}[c][][1]{$x$ (mm)}
\psfrag{S}[c][][1]{$\sigma_o/2\Omega$}
\psfrag{W}[c][][1]{$\omega_{max}$ (rad~s$^{-1}$)}
\psfrag{H}[c][][1]{$g(\theta)$}
\psfrag{EQ2}[r][][0.8]{}
\begin{center}
\centerline{\includegraphics[width=7cm]{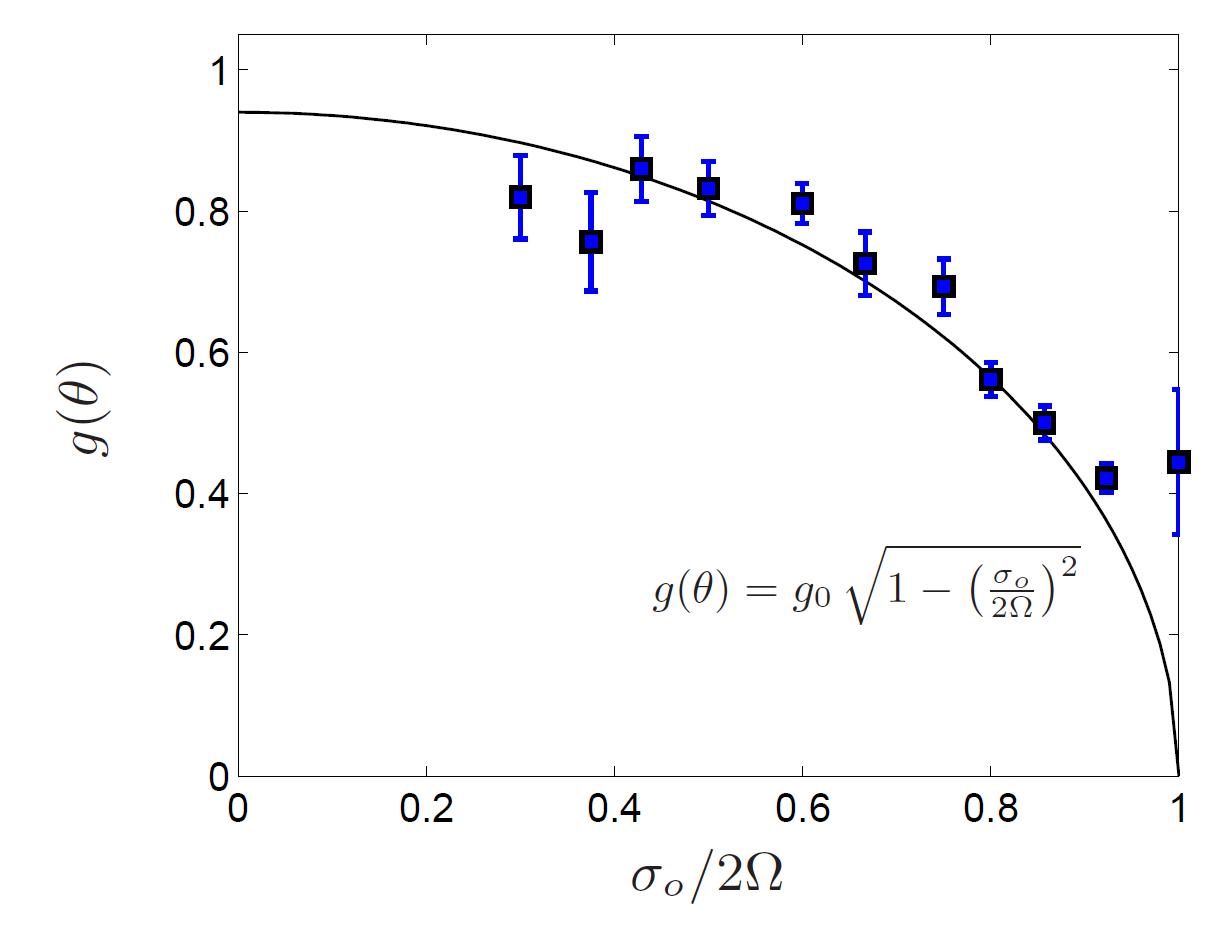}}
\caption{Forcing efficiency $g(\theta)$ defined from
Eq.~(\ref{eq:defg}) as a function of $\sigma_o / 2\Omega$. Squares
and errorbars represents respectively the mean and the standard
deviation for each $\sigma_o / 2\Omega$ ratio (reflecting the
variability of $\omega_{max}$ along $x$). The black line shows a
best fit according to Eq.~(\ref{eq:gm}), with $g_0 =0.94\pm
0.10$.} \label{fig:vortampb}
\end{center}
\end{figure}

The vorticity scale $W_0^*$ is theoretically related to the
velocity scale $U_0^*$ through the relation $W_0^* = (E_1(0) /
E_0(0)) U_0^* / \ell \simeq 0.506\,U_0^*/\ell$ (see the Appendix).
Since the wavemaker velocity is $\sigma_o A$, the velocity scale
$U_0^*$ is expected to write in the form $\sigma_o A \,
g(\theta)$, where the unknown function $g(\theta)$ describes the
forcing efficiency of the wavemaker. Accordingly, the forcing
efficiency can be deduced from the vorticity data, by computing
\begin{equation}
g(\theta) = \frac{W_0^*}{0.506\,\sigma_o A / \ell} =
\frac{\omega_{max} \,(x/\ell)^{2/3}}{0.506\,\sigma_o A / \ell}
\label{eq:defg}
\end{equation}
for each value of $\sigma_o / 2\Omega$. Measurements of
$g(\theta)$ are plotted as a function of $\sigma_o / 2\Omega$ in
Fig.~\ref{fig:vortampb}. As expected, this forcing efficiency
decreases as $\sigma_o/2\Omega$ is increased, i.e. as the wave
beam becomes closer to the horizontal. In the limit $\sigma_o /
2\Omega \rightarrow 1$, the vertically oscillating wavemaker
becomes indeed very inefficient to force the quasi-horizontal
velocities of the wave.

An analytical expression for the function $g(\theta)$ would
require to solve exactly the velocity field in the vicinity of the
wavemaker, and in particular the coupling between the oscillating
boundary layer and the wave far from the source, which is beyond
the scope of this paper. In the case of a cylinder, a naive
estimate of $g(\theta)$ could however be obtained, assuming that
the effective velocity forcing is simply given by the
projection of the wavemaker velocity along the wave beam
direction, yielding
\begin{equation}
g(\theta) = g_0 \sin \theta = g_0 \,
\sqrt{1-\left(\frac{\sigma_o}{2\Omega}\right)^2}, \label{eq:gm}
\end{equation}
with $g_0$ a constant to be determined. A best fit of the
experimental values of $g(\theta)$ with this law leads to $g_0
\simeq 0.94 \pm 0.10$ [see Fig.~\ref{fig:vortampb}], and
reproduces well the decrease of $g(\theta)$ as $\sigma_o/2\Omega$
is increased. The fact that $g_0$ is found close to 1 indicates that the
inertial wave beam is essentially fed by the oscillating
velocity field in the close vicinity of the wavemaker.
The discrepancy at large forcing frequency may be
due to the breakdown of the similarity solution as the angle
$\theta$ approaches $0$.

\section{Stokes drift}

We finally consider the possibility of a Stokes drift along the
wavemaker which is expected because of the attenuation of the wave
amplitude induced by the viscosity. A fluid particle in the
inertial wave approximately describes a circular orbit. During
this orbit, the particle experiences a larger velocity along $y$
when it is closer than when it is further from the wavemaker (see
Fig. \ref{fig:frame}), resulting in a net mass transport along
$y$. This is similar to the classical Stokes drift for water
waves, which is horizontal because of the decay of the velocity
magnitude with depth.\cite{Longuet1953} Here the drift is expected
in general in the direction given by $\boldsymbol \Omega \times
{\bf c}_g$, since the viscous attenuation takes place along the
group velocity ${\bf c}_g$.

Attempts to detect this effect have been carried out, from PIV
measurements in vertical planes $(Y,Z)$. Because of the weakness
of the considered drift, the measurements have been performed very
close to the wavemaker, for $X$ between 5 and 30~mm, where a
stronger effect is expected. However, those attempts were not
successful, probably because the drift, if present, is hidden by
the stronger fluid motions induced by the residual thermal
convection columns, as discussed in Sec.~\ref{sec:piv}.

The magnitude of the expected Stokes drift cannot be easily
inferred from the complex motion of the fluid particles close to
the wavemaker. An estimate could however be obtained in the far
field, from the similarity solution of the wave beam. We consider
for simplicity a particle lying at the center of the wave beam ($z
= 0$), at a mean distance $x_0$ from the source, describing
approximate circles of gyration radius $a \simeq | {\bf u}(x_0) |
/ \sigma$ in the tilted plan $(x,y)$. The expected drift velocity
$\overline{\bf v}_S$ can be approximated by computing the velocity
difference between the two extreme points $x_0 - a$ and $x_0 + a$
of the orbit, yielding, to first order in $a/x_0$
to,\cite{coeffdrift}
\begin{equation}
\overline{v}_{Sy} (x_0) \simeq \frac{2}{3} \frac{U_0^{*2}
\ell^{2/3}}{\sigma x_0^{5/3}}. \label{eq:sd}
\end{equation}
The steep decrease as $x_0^{-5/3}$ confirms that the drift should
be essentially present close to the wavemaker. Although this
formula is expected to apply only in the far-field wave (typically
for $x_0 > 40 \ell$, see the Appendix), its extrapolation close to
the wavemaker, for $x_0 \simeq 10 \ell \simeq 2 R$, gives
$\overline{v}_{Sy} \simeq 0.1$~mm~s$^{-1}$. This expected drift
velocity is about 10\% of the wave velocity at the same location,
but it turns out to remain smaller than the velocity contamination
due to the thermal convection columns. Although the phase
averaging proved to be efficient to extract the inertial wave
field from the measured velocity field --~thanks to a frequency
separation between convection effects and inertial wave~--, it
fails here to extract the much weaker velocity signal expected
from this drift, since it is zero frequency and hence mixed with
the very low frequency convective noise.

\section{Conclusion}

In this paper, Particle Image Velocimetry measurements have been
used to provide quantitative insight into the structure of the
inertial wave emitted by a vertically oscillating horizontal
cylinder in a rotating fluid. Large vertical fields of view could
be achieved thanks to a new rotating platform, allowing for direct
visualization of the cross-shaped {\it St. Andrew's} wave pattern.

It must be noted that performing accurate PIV measurements of the
very weak signal of an inertial wave is a challenging task. In
spite of the high stability of the angular velocity of the
platform ($\Delta\Omega/\Omega < 5\,10^{-4}$), the velocity
signal-to-noise remains moderate here. Additionally, slowly
drifting vertical columns are present because of residual thermal
convection effects, and are found to account for most of velocity
noise in these experiments. Those thermal convection effects are
very difficult to avoid in large containers, even in an
approximately thermalized room. However, this noise can be
significantly reduced by a phase-averaging over a large number of
oscillation periods. This concern is not present for internal
waves in stratified fluids, because residual thermal motions are
inhibited by the stable stratification. This emphasizes the
intrinsic difficulty of experimental investigation of inertial
waves, in contrast to internal waves which have been the subject
of a number of studies.

In this article, emphasis has been given on the spreading of the
inertial wave beam induced by viscous dissipation. The attenuation of a
two-dimensional  wave beam emitted from a linear source is
purely viscous, whereas it combines viscous and geometrical
effects in the case of a conical wave emitted from a point source.
The linear theory presented in this paper is derived under the
classical boundary-layer assumptions first introduced by Thomas
and Stevenson\cite{Thomas1972} for two-dimensional internal waves
in stratified fluids. The measured thickening of the wave beam and
the decay of the vorticity envelope are quantitatively fitted by
the scaling laws of the similarity solutions of this linear
theory, $\delta(x) \sim x^{1/3}$ and $\omega_{max}(x) \sim
x^{-2/3}$, where $x$ is the distance from the source. More
precisely, we have shown that the amplitude of the vorticity
envelope could be correctly predicted from the velocity
disturbance induced by the wavemaker, by introducing a simple
forcing efficiency function $g(\theta)$, where $\theta$ is the
angle of the wave beam.

Finally, it is shown that an attenuated inertial wave beam should
in principle generate a Stokes drift along the wavemaker, in the
direction given by $\boldsymbol \Omega \times {\bf c}_g$, where
${\bf c}_g$ is the group velocity. However, in spite of the high
precision of the rotating platform and the PIV measurements,
attempts to detect this drift were not successful in the present
configuration. Velocity fluctuations induced by thermal convection
effects probably hide this slight mean drift velocity, suggesting
that an improved experiment with a very carefully controlled
temperature stability would be necessary to detect this very weak
effect.

\acknowledgments

We acknowledge A. Aubertin, L. Auffray, C. Borget, G.-J. Michon
and R. Pidoux for experimental help, and T. Dauxois, L. Gostiaux,
M. Mercier, C. Morize, M. Rabaud and B. Voisin for fruitful
discussions.  The new rotating platform ``Gyroflow'' was funded by
the ANR (grant no. 06-BLAN-0363-01 ``HiSpeedPIV''), and the
``Triangle de la Physique''.

\appendix

\section{Similarity solution for a viscous planar inertial wave}

In this appendix, we derive the similarity solution for a viscous
planar inertial wave, following the procedure first described by
Thomas and Stevenson\cite{Thomas1972} for internal waves.

We consider the inertial wave emitted from a thin linear
disturbance invariant along the $Y$ axis and oscillating along $Z$
with a pulsation $\sigma$ in a viscous fluid rotating at angular
velocity ${\boldsymbol \Omega} = \Omega {\bf e}_Z$. Since the
linear source is invariant along $Y$, so will the wave beams, and
the energy propagates in the $(X,Z)$ plan. In the following, we
consider only the wave beam propagating along $X>0$ and $Z>0$.

The linearized vorticity equation is
$$
\partial_t {\boldsymbol \omega} = (2 {\boldsymbol \Omega} \cdot \nabla) {\bf u} + \nu \nabla^2 {\boldsymbol \omega}.
$$
Recasting the problem in the tilted frame of the wave, $({\bf
e}_x, {\bf e}_y, {\bf e}_z)$, with ${\bf e}_y = {\bf e}_Y$, and
${\bf e}_x$ tilted of an angle $\theta = \cos^{-1} (\sigma /
2\Omega)$ with the horizontal, one has ${\boldsymbol \Omega} =
\Omega\,(\sin \theta \,{\bf e}_x + \cos \theta \,{\bf e}_z)$, so
that $(2 {\boldsymbol \Omega} \cdot \nabla) = 2 \Omega \,(\sin
\theta \,\partial_x + \cos \theta \, \partial_z) = \sigma\,(\tan
\theta\, \partial_x + \partial_z)$. Assuming that the flow inside
the wave beam is quasi-parallel (boundary layer approximation),
i.e., such that $|u_x|, |u_y| \gg |u_z|$, $|\omega_x|, |\omega_y|
\gg |\omega_z|$, and $\nabla^2 \simeq \partial_z^2$, the
linearized vorticity equation reduces to
\begin{eqnarray}
\partial_t \omega_x &=& \sigma (\tan \theta \partial_x + \partial_z) u_x + \nu \partial_z^2 \omega_x, \label{eq:wx} \\
\partial_t \omega_y &=& \sigma (\tan \theta \partial_x + \partial_z) u_y + \nu \partial_z^2 \omega_y. \label{eq:wy}
\end{eqnarray}
We introduce the complex velocity and vorticity fields in
the $(x,y)$ plan as
$$
U = u_x + i u_y, \qquad W = \omega_x + i \omega_y.
$$
Since, within the quasi-parallel approximation, one has $W = i
\partial_z U$, the combination (\ref{eq:wx})$+i$(\ref{eq:wy})
yields
\begin{equation}
i \partial_t \partial_z U = \sigma (\tan \theta \partial_x +
\partial_z) U + i \nu \partial_z^3 U. \label{eq:a}
\end{equation}
Searching solutions in the form $U = U_0 e^{-i \sigma t}$,
Eq.~(\ref{eq:a}) becomes
\begin{equation}
\partial_x U_0 + i \ell^2 \partial_z^3 U_0 = 0,
\label{eq:b}
\end{equation}
where we have introduced the viscous scale $\ell$ (\ref{eq:ell}).
Eq.~(\ref{eq:b}) admits similarity solutions as a function of the
variable:
\begin{equation}
\eta = \frac{z}{x^{1/3} \ell^{2/3}}, \label{eq:u0ante}
\end{equation}
which are of the form:
\begin{equation}
U_0 (x,z) = \tilde U_0 \left( \frac{\ell}{x} \right)^{1/3}
f(\eta), \label{eq:u0a}
\end{equation}
and where $\tilde U_0$ is a velocity scale and $f(\eta)$ a
non-dimensional complex function of the reduced transverse
coordinate $\eta$. Plugging such similarity solution
(\ref{eq:u0a}) into Eq.~(\ref{eq:b}) shows that $f(\eta)$ is
solution of the ordinary differential equation
\begin{equation}
3 f''' + i(f + \eta f') = 0, \label{eq:f}
\end{equation}
which is identical to the equation (16) derived by Thomas and
Stevenson\cite{Thomas1972} for the pressure field of internal
waves. Following their development, we introduce the family of
functions $f_m$ defined through
\begin{equation}
f_m(\eta) = c_m + i s_m = \int_0^{\infty} K^m e^{-K^3} e^{iK \eta}
dK, \label{eq:cmsm}
\end{equation}
where $c_m$ and $s_m$ are real, and such that $f_0(\eta)$ is a
solution of Eq.~(\ref{eq:f}).

The velocity in the plan of the wave beam is therefore given by $u_x =
\Re \{ U \}$ and $u_y = \Im \{ U \}$, leading to
\begin{eqnarray}
u_x = \frac{U_0^*}{E_0(0)} \left( \frac{\ell}{x} \right)^{1/3} \left[c_0(\eta) \cos (\sigma t) + s_0(\eta) \sin (\sigma t)\right] \nonumber \\
u_y = \frac{U_0^*}{E_0(0)} \left( \frac{\ell}{x} \right)^{1/3}
\left[s_0(\eta) \cos (\sigma t) - c_0(\eta) \sin (\sigma t)\right]
\nonumber
\end{eqnarray}
with $U^*_0 = E_0(0) \tilde U_0 \simeq 0.893 \tilde U_0$ where we
introduce the family of envelopes $E_m (\eta) = |f_m(\eta)| =
(c_m^2 + s_m^2)^{1/2}$ for $m=0,1$.

Similarly, the vorticity in the plan of the wave beam is $\omega_x
= \Re \{ W \}$ and $\omega_y = \Im \{ W \}$, so that
\begin{eqnarray}
\omega_x = \frac{W_0^*}{E_1(0)} \left( \frac{\ell}{x} \right)^{2/3} \left[-c_1(\eta) \cos (\sigma t) - s_1(\eta) \sin (\sigma t)\right], \nonumber \\
\omega_y = \frac{W_0^*}{E_1(0)} \left( \frac{\ell}{x}
\right)^{2/3} \left[-s_1(\eta) \cos (\sigma t) + c_1(\eta) \sin
(\sigma t)\right], \nonumber
\end{eqnarray}
with $W_0^* = \left[E_1(0) / E_0(0)\right] U_0^* / \ell \simeq
0.506 U_0^* / \ell$.

\begin{figure}
\psfrag{ETA}[c][][1.1]{$\eta$}
\psfrag{E0E1}[c][][1.1]{$E_{m}(\eta)/E_{m}(0)$}
\begin{center}
\vspace{0.6cm}
\centerline{\includegraphics[width=6.5cm]{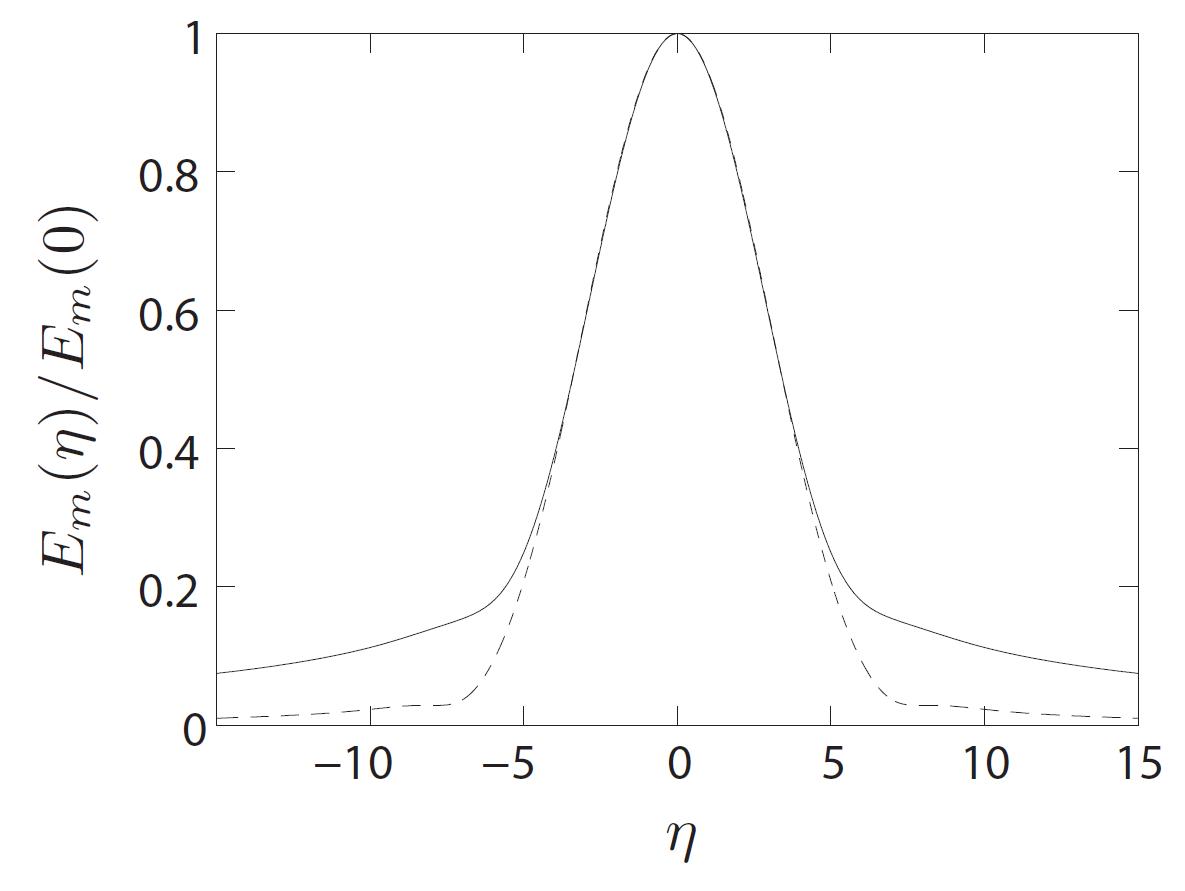}}
\caption{Normalized velocity (---, $m=0$) and vorticity (- -,
$m=1$) envelopes of the similarity solutions.} \label{fig:e0e1}
\end{center}
\end{figure}

The velocity and vorticity envelopes, defined as $u_{0} = (
\langle u_x^2 \rangle + \langle u_y^2 \rangle)^{1/2}$ and
$\omega_{0} = (\langle \omega_x^2 \rangle + \langle \omega_y^2
\rangle)^{1/2}$, where $\langle \cdot \rangle$ is the time-average
over one wave period, are given by
\begin{eqnarray}
u_{0} = U_0^* \left( \frac{\ell}{x} \right)^{1/3} \frac{E_0 (\eta)}{E_0 (0)}, \nonumber \\
\omega_{0} = W_0^* \left( \frac{\ell}{x} \right)^{2/3} \frac{E_1
(\eta)}{E_1 (0)}. \nonumber
\end{eqnarray}
The two normalized envelopes $E_m(\eta) / E_m(0)$ are compared in
Fig.~\ref{fig:e0e1}. Interestingly, they closely
coincide up to $\eta \simeq 4$, but the vorticity envelope
decreases much more rapidly than the velocity envelope as $\eta
\rightarrow \infty$ (one has $E_m \propto 1/\eta^{m+1}$ for $\eta
\gg 1$).

It is interesting to note that velocity and vorticity in the
present analysis are analogous to the pressure and velocity in the
analysis of Thomas and Stevenson.\cite{Thomas1972} One consequence
is that the lateral decay of the velocity envelope is sharper for
an internal wave (as $1/\eta^2$) than for an inertial wave (as
$1/\eta$).

Finally, the $z$ component of the velocity is obtained using
incompressibility ($\partial_x u_x + \partial_z u_z = 0$), yielding
\begin{equation}
u_z = \frac{1}{3} u_x \eta \left( \frac{\ell}{x} \right)^{2/3} =
\frac{1}{3} u_x \frac{z}{x}.
\end{equation}
This result shows that the streamlines projected in the $(x,z)$
plan are along lines of constant phase (i.e., constant $\eta$). As
a consequence, a particle trajectory is an approximate circle,
projected on curved surfaces, invariant along $y$, and such that
$z = \eta^* \ell^{2/3} x^{1/3}$, with $\eta^*$ given by the
initial location.

The thickness $\eta_{1/2}$ of the velocity and vorticity envelopes
are defined  such that $E_m(\eta_{1/2}/2) = E_m(0)/2$. Those
thicknesses turn out to be almost equal: $\eta_{1/2} \simeq 6.841$
for $m=0$ and $\eta_{1/2} \simeq 6.834$ for $m=1$. In dimensional
units, the wave thickness is thus given by Eq.~(\ref{eq:delta}).
Interestingly, the envelope of $u_z$ is given by $\eta E_0(\eta)$,
which tends towards $1$ as $\eta \rightarrow \infty$, so that no
thickness could be defined for $u_z$.

Finally, we note that the quasi-parallel approximation used in the
present analysis is satisfied for $|u_z| / |u_x| \ll 1$. Using
$|u_z| / |u_x| = \eta^{1/3}\,(\ell / x)^{1/3} / 3$, and evaluating
the envelope ratio at the boundary of the wave, i.e., for $\eta =
\eta_{1/2} / 2 \simeq 3.42$, this criterion is satisfied within
10\% for $x > 38 \ell$.


\begin{thebibliography}{}
\bibitem{coeffdrift}
A numerical integration of the particle trajectory in the center
of the beam ($z=0$) actually shows that the numerical prefactor in Eq.~(\ref{eq:sd}) is 1.047 instead of 2/3.
\end{thebibliography}
\end{document}